\documentclass[3p,authoryear]{elsarticle}
\usepackage{epstopdf}
\usepackage{subfigure,graphicx}
\usepackage{float}
\usepackage{amsmath, amsfonts, amssymb,mathrsfs}
\usepackage{amsthm}
\usepackage{epsf}
\usepackage{amsfonts}
\usepackage{stmaryrd}
\usepackage{amssymb}
\usepackage{leftidx}
\usepackage{xcolor}
\usepackage{mathtools}
\usepackage{placeins}
\usepackage{booktabs}
\usepackage{enumitem}
\usepackage{caption}
\usepackage{multirow}
\usepackage{stackengine}
\usepackage[utf8]{inputenc}
\usepackage[english]{babel}
\usepackage{color}
\usepackage{hyperref}

\theoremstyle{plain}

\def\bfu{{\bf u}}
\def\bfv{{\bf v}}

\def\bfE{{\bf E}}

\def\bfI{{\bf I}}

\def\bfN{{\bf N}}

\def\bfS{{\bf S}}

\def\bfX{{\bf X}}

\newcommand\sts{s_{\texttt{ts}}}
\newcommand\scs{s_{\texttt{cs}}}
\newcommand\shs{s_{\texttt{hs}}}
\newcommand\sbs{s_{\texttt{bs}}}
\newcommand\sss{s_{\texttt{ss}}}
\newcommand\e{\varepsilon}

\long\def\symbolfootnote[#1]#2{\begingroup%
\def\thefootnote{\fnsymbol{footnote}}\footnote[#1]{#2}\endgroup}

\makeatletter
\renewcommand\@biblabel[1]{}

\makeatother

\begin{document}
\begin{frontmatter}

\title{Classical variational phase-field models cannot predict fracture nucleation\vspace{-0.2cm}}

\author{Oscar Lopez-Pamies}
\ead{pamies@illinois.edu}

\author{John E. Dolbow}
\ead{jdolbow@duke.edu}

\author{Gilles A. Francfort}
\ead{gfrancfort@flatironinstitute.org}

\author{Christopher J. Larsen}
\ead{cjlarsen@wpi.edu}

\vspace{-0.2cm}

\address{Department of Civil and Environmental Engineering, University of Illinois, Urbana--Champaign, IL 61801, USA  \vspace{-0.1cm}}

\address{Department of Mechanical Engineering, Duke University, Durham, NC 27708, USA \vspace{-0.1cm}}

\address{Flatiron Institute, 162 Fifth Avenue, New York, NY 10010, USA \vspace{-0.1cm}}

\address{Department of Mathematical Sciences, Worcester Polytechnic Institute, Worcester, MA 01609-2280, USA \vspace{-0.1cm}}

\vspace{0.0cm}

\begin{abstract}

\vspace{0.0cm}

Notwithstanding the evidence against them, classical variational phase-field models continue to be used and pursued in an attempt to describe fracture nucleation in elastic brittle materials. In this context, the main objective of this paper is to provide a comprehensive review of the existing evidence against such a class of models as descriptors of fracture nucleation. To that end, a review is first given of the plethora of experimental observations of fracture nucleation in nominally elastic brittle materials under quasi-static loading conditions, as well as of classical variational phase-field models, without and with energy splits. These models are then confronted with the experimental observations. The conclusion is that they cannot possibly describe fracture nucleation in general. This because classical variational phase-field models cannot account for material strength as an independent macroscopic material property. The last part of the paper includes a brief summary of a class of phase-field models that can describe fracture nucleation. It also provides a discussion of how pervasively material strength has been overlooked in the analysis of fracture at large, as well as an outlook into the modeling of fracture nucleation beyond the basic setting of elastic brittle materials.  

\vspace{0.0cm}

\endkeyword

\end{abstract}

\end{frontmatter}

\vspace{-0.2cm}

\section{Introduction}\label{Sec: Intro}

Historically, research on fracture has primarily focused on the growth of pre-existing cracks, or crack propagation, while much less effort has been devoted to the nucleation of cracks. Since the early 2010s, however, the subject of crack nucleation has become the focus of an increasing number of investigations, especially within the basic setting of elastic brittle materials\footnote{Brittleness, as we understand it, is the assumption that the energy dissipated through a crack (add)-surface is proportional to that add-surface, the coefficient of proportionality being called the fracture toughness, or for reasons that are familiar to the practitioner of fracture, the critical energy release rate. Elastic brittle materials are thus materials that, in response to mechanical forces, either deform elastically or fracture in a brittle manner, the latter being the sole mechanism by which these materials dissipate energy.} and quasi-static loading conditions. The next paragraphs trace back the origin of this recent impetus to two disparate events: the popularization of phase-field approximations of the variational theory of brittle fracture and the uncovering of the so-called phenomenon of cavitation in elastomers as a fracture event.

The origin of phase-field approximations of the variational sharp theory of brittle fracture \citep{Francfort98} --- dubbed here as \emph{classical variational phase-field models} or simply \emph{variational phase-field models} for short --- dates back to the turn of the millennium \citep{Bourdin00}.\footnote{Throughout, by variational phase-field models we mean phase-field models of fracture that $\Gamma$-converge to the variational theory of brittle fracture of \cite{Francfort98}; see Section \ref{Sec: The P-F model} below. Many other phase-field models of fracture are variational but do not $\Gamma$-converge to that theory; see, e.g., \cite{Lorentz11}, \cite{Iurlano16}, \cite{Wu17}, and \cite{LDLP24}.} In spite of the pioneering efforts of Blaise Bourdin who undertook the non-trivial numerical implementation of phase-field evolution within that context, it took the computational mechanics community a while to become receptive to  the approach. It did so about a decade later, when it became extensively used; see, e.g., \cite{Amor09,Miehe10,Muller10,Landis12}. With the increasing popularity came the realization that these models could  describe fracture nucleation in a way that the variational theory of sharp fracture cannot. This quirk --- described by many as ``getting fracture nucleation for free'' --- comes about because of the use of a small but finite regularization length $\varepsilon$ for crack thickness. This observation soon resulted in a reimagining of the regularization length $\varepsilon$ in variational phase-field models of fracture as a material length scale \citep{Freddi10,Marigo11}, of course at the expense of severing the connection with the variational theory of brittle fracture, or with any other theory of sharp fracture for that matter. Comfort with this guiding principle increased with apparent experimental agreement \citep{Tanne18} and models with an $\varepsilon$ slaved to some material-specific value were and continue to be used and pursued in the literature in an attempt to describe fracture nucleation in elastic brittle materials.

In a different place, cavitation in elastomers had, since the 1950s, been viewed as a purely elastic process.\footnote{For an account of the fascinating history of cavitation in elastomers, see, e.g., the recent review by \cite{BCLLP24}.} As such it lent itself to elegant mathematics first expounded in \cite{Ball82} and subsequently indulged in many works. Yet, despite the many investigations of cavitation in elastomers as a purely elastic process for decades, it was only in 2015 that direct comparisons between the elasticity view of this phenomenon and experiments were reported in the literature \citep{LRLP15}. Those  revealed that cavitation in elastomers is in fact a fracture nucleation event, one that could not be explained by any existing theory at the time, including existing variational phase-field models. 

The question before us was clear, although largely unstated at the time: What are the intrinsic macroscopic material properties that govern fracture nucleation? It was posed as such in \cite*{KFLP18} and \cite{KLP20} for nominally nonlinear elastic brittle materials (e.g., elastomers like silicone) and in \cite{KBFLP20} for nominally linear elastic brittle materials (e.g., ceramics like titania, alumina, graphite, and polymers like PMMA); see also \cite{LP23}. 

The answer, based on a vast body of experimental evidence gathered for over a century, is that at least three intrinsic macroscopic material properties govern fracture nucleation in elastic brittle materials:
\begin{enumerate}[label=\Roman*.]

\item{The elasticity of the material;}

\item{Its strength; and}

\item{Its intrinsic fracture toughness or critical energy release rate.}

\end{enumerate}
Out of these, the strength has been the most often misunderstood, insomuch that no precise complete definition appears in the classical literature. \cite{KLP20} and \cite{KBFLP20} proposed the following definition: the strength of an elastic brittle material is the set of all critical stresses $\bfS$ at which the material fractures when it is subjected to a state of monotonically increasing, spatially uniform, but otherwise arbitrary stress. Such a set of critical stresses defines a surface $\mathcal{F}(\bfS)=0$ in stress space, which is referred to as the \emph{strength surface} of the material. 

Like elasticity and toughness, strength is an intrinsic macroscopic material property that can be measured directly from macroscopic experiments. Critically, while they may correlate with some of the same underlying microscopic features of the material, the elasticity, the strength, and the toughness are independent of each other. What is more, save for some elementary restrictions that they must satisfy,\footnote{For instance, in isotropic materials, the Young's modulus $E$ and Poisson's ratio $\nu$ must satisfy $E>0$ and $-1\leq\nu\leq 1/2$, but are otherwise unconstrained. Similarly, the toughness $G_c$ is positive but  otherwise arbitrary.} they can take on any value. So the strength surface $\mathcal{F}(\bfS)=0$ is potentially any star-shaped surface in stress space containing $0$ in the interior of the ascribed domain.\footnote{Accordingly, rays starting at the origin $\bfS=0$ can cross the strength surface $\mathcal{F}(\bfS)=0$ at most once.}

In view of the above realization, \cite*{KFLP18}, \cite{KLP20}, and \cite{KBFLP20} established that any potentially successful attempt at a macroscopic theoretical description of fracture nucleation in elastic brittle materials must be able to account for any elasticity, any strength, and any toughness. As a corollary, no variational phase-field model can possibly describe fracture nucleation in general. This is because such models inherently only account for the elasticity and the toughness of the material, but not for its strength. Sure, as shown in Section 3 of \cite{KBFLP20}, these models can be made to account for certain types of strength surfaces upon an appropriate choice of regularization length $\varepsilon$ and energy split. However, the strength surfaces that result from such an approach are not representative of actual materials. The reason for this shortcoming is that those strength surfaces are subordinate to the elastic energy and the toughness of the material, instead of being truly independent macroscopic material properties, as observed experimentally.

Notwithstanding the results just outlined, variational phase-field models have continued to be widely used by the computational mechanics community, presumably because they are typically employed in particular problems where only a small part of the strength surface --- the one that is fitted to some desired outcome by a suitable choice of the value of $\varepsilon$ --- is relevant and hence their inability to model fracture nucleation in general is not apparent; see, e.g., \cite{Paneda18,Kaliske19,WKLiu19,Bouklas20,Wriggers20,Reddy20,Keip21,Rabczuk22,Shen23,Linder24}. What is more, faced with a precise definition of strength and with the realization that different energy splits lead to different strength surfaces, researchers have continued to pursue different types of energy splits in the hope of producing variational phase-field models that better approximate the strength surfaces $\mathcal{F}(\bfS)=0$ of actual materials; see, e.g., \cite{Maurini22} and \cite{Maurini24}. 

In this context, the main objective of this paper is to provide a comprehensive review of the existing evidence that conclusively establishes that variational phase-field models cannot be valid descriptors of fracture nucleation. We begin in Section \ref{Sec: Experiments} by providing a summary of the current experimental knowledge of fracture nucleation in nominally elastic brittle materials under quasi-static loading conditions. In Section \ref{Sec: The P-F model}, we introduce the variational phase-field models, without and with energy splits. In Section \ref{Sec: The evidence}, we then confront the models with basic experimental observations on a variety of materials of common use (graphite, titania, natural rubber, and a synthetic rubber), thereby demonstrating that variational phase-field models cannot possibly describe fracture nucleation precisely because they can only account for strength surfaces that are written in terms of the elastic energy and the toughness of the material. As already stated, strength surfaces of actual materials are independent macroscopic material properties! We conclude in Section \ref{Sec: Final Comments} by providing a brief summary of a class of phase-field models that can describe fracture nucleation, that introduced by \cite*{KFLP18}. We also provide an outlook into the modeling of fracture nucleation beyond the basic setting of elastic brittle materials.

\section{A summary of experimental observations of fracture nucleation in nominally elastic brittle materials}\label{Sec: Experiments}

Experimental investigations of the nucleation of fracture in nominally elastic brittle materials are broadly of three distinct sorts: when the state of stress is spatially uniform, when a large pre-existing crack is present, and if neither holds true. The findings can be summarized as follows.

\subsection{Nucleation under states of spatially uniform stress: The strength}\label{Sec: Strength ingredient}

A plethora of experiments carried out since the 1830s \citep{Lame1833} have repeatedly shown that when a macroscopic piece of a nominally elastic brittle material is subjected to a state of monotonically increasing, spatially uniform, but otherwise arbitrary stress, fracture will nucleate at a critical value of the applied stress. The set of all such critical stresses defines a surface in stress space. This surface is referred to as the strength surface of the material. In terms of the first Piola-Kirchhoff stress tensor $\bfS$, we write
\begin{equation}
\mathcal{F}(\bfS)=0. \label{FPK}
\end{equation}
Any other stress measure could be equally used, but some prove more convenient than others \citep{KLP20}. We will agree that any stress state $\bfS$ such that 
\begin{equation}
\mathcal{F}(\bfS)\geq0 \label{FPK-bound}
\end{equation}
is in violation of the strength of the material.

Physically, the origin of the type of fracture nucleation described by the strength surface (\ref{FPK}) can be directly linked to the underlying defects of the material at hand. Put differently,   the strength surface (\ref{FPK}) is the macroscopic manifestation of the presence of microscopic defects, i.e., the ``weakest links'' in the material. This stands in stark contrast with  other macroscopic properties impervious to microscopic defects, such as the elasticity. The nature of the defects is material dependent. For instance, they can be pores at grain boundaries in a sintered ceramic \citep{Kovar00} or inhomogeneous distributions of cross-links in an elastomer \citep{Valentinetal2010}. Furthermore, the size and spatial variations of defects are inherently stochastic. The variations are most acute when comparing material points within the bulk of the body with material points on its boundary, since different fabrication processes or boundary treatments (such as polishing or chemical treatments) can drastically affect the nature of boundary defects vis-\`a-vis those in the bulk. It is for this reason that the strength surface (\ref{FPK}) is a material property that is inherently stochastic.

While experiments that measure the uniaxial tensile strength $\sts$ --- that is, the point defined by the equation $\mathcal{F}({\rm diag}(\sts>0, 0, 0)) = 0$ --- are, in general, fairly accessible, direct experiments that probe triaxial states of stress are much less so.

\begin{figure}[t!]
   \centering \includegraphics[width=0.9\linewidth]{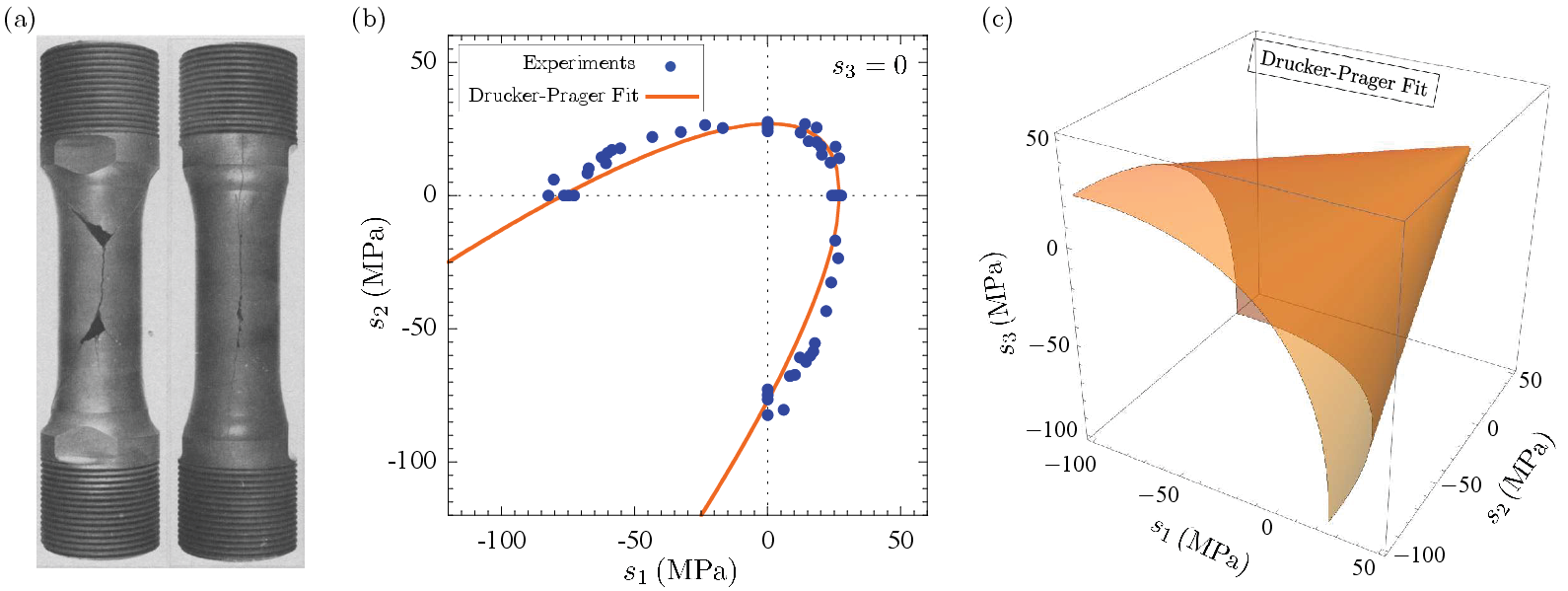}
   \caption{\small The experiments of \cite{Sato87} to measure the strength of IG-11 graphite. (a) Typical nucleated cracks in thin-walled tubes subjected to a combination of pressurization and axial loading. (b) Plot of the principal stress $s_2$ in terms of the principal stress $s_1$ at fracture nucleation; the results correspond to the case when roughly $s_3=0$. (c) Plot of the Drucker-Prager strength surface (\ref{DP}), fitted to the experimental data, in the space of all three principal stresses $(s_1,s_2,s_3)$.}
   \label{Fig1}
\end{figure}
For hard materials, probing triaxial stress states of the  form $\bfS={\rm diag}(s_1,s_2,0)$ has been achieved by using thin-walled tubes subjected to internal and external pressurization at the same time that they are axially loaded. By way of an example, Fig.~\ref{Fig1} reproduces the strength results for IG-11 graphite obtained by \cite{Sato87} using such a test. The plots in Fig.~\ref{Fig1} include the fit of the experimental data (solid circles) by the Drucker-Prager strength surface 
\begin{equation}
\mathcal{F}(\bfS)=\sqrt{\mathcal{J}_2}+\dfrac{\scs-\sts}
{\sqrt{3}\left(\scs+\sts\right)}\, \mathcal{I}_1-\dfrac{2\scs\sts}
{\sqrt{3}\left(\scs+\sts\right)}=0\qquad {\rm with}\qquad \left\{\hspace{-0.1cm}\begin{array}{l}\sts=27\,{\rm MPa}\vspace{0.2cm}\\
\scs=77\,{\rm MPa}\end{array}\right., \label{DP}
\end{equation}
where
\begin{equation}\label{Inv-S}
\mathcal{I}_1:=s_1+s_2+s_3\qquad {\rm and}\qquad  \mathcal{J}_2:=\dfrac{1}{3}\left((s_1+s_2+s_3)^2-s_1^2-s_2^2-s_3^2\right),
\end{equation}
so to reinforce the crucial realization that the strength of a material is characterized by an entire surface in stress space, and not by just a single point in that space, a common a priori in the literature. In the above expressions, $s_1$, $s_2$, $s_3$ stand for the eigenvalues of the Biot stress tensor, or principal nominal stresses, while the material constants $\sts$ and $\scs$ denote the uniaxial tensile and compressive strengths of the material, that is, they denote the critical nominal stress values at which fracture nucleates under spatially uniform states of monotonically increased uniaxial tension and compression when $\bfS={\rm diag}(s>0,0,0)$ and $\bfS={\rm diag}(s<0,0,0)$, respectively.\footnote{For later use, we introduce here the analogous notation $\sbs$, $\shs$, and $\sss$ for the equi-biaxial tensile, hydrostatic tensile, and shear strengths, that is, the critical values of the nominal stress $s$ at which fracture nucleates when $\bfS={\rm diag}(s>0,s>0,0)$, $\bfS={\rm diag}(s>0,s>0,s>0)$, and $\bfS={\rm diag}(s>0,-s,0)$, respectively.}

\begin{figure}[t!]
   \centering \includegraphics[width=0.9\linewidth]{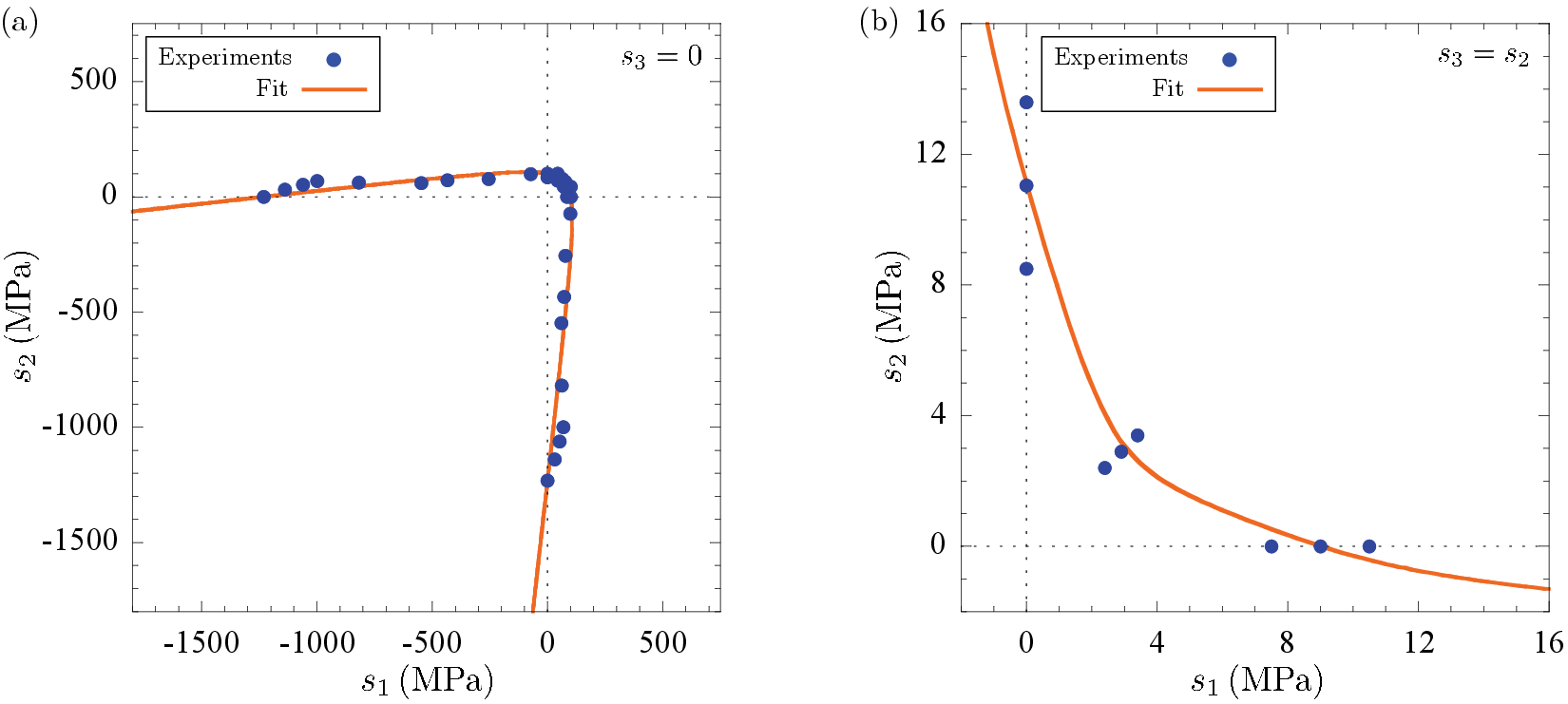}
   \caption{\small (a) Plot of the strength surface for titania reported by \cite{Ely72}. The data is plotted in the space of principal stresses $(s_1, s_2)$ with $s_3 = 0$. (b) Plot of the strength surface for one of the natural rubbers studied by \cite{GL59}. The data is plotted in the space of principal stresses $(s_1, s_2)$ with $s_3 = s_2$.}
   \label{Fig2}
\end{figure}

For soft materials, the strength under triaxial stress states of the form $\bfS={\rm diag}(s_1,s_2,0)$ has been traditionally measured by biaxial and pure-shear stretching of thin sheets \citep{Kawabata73}, while the so-called poker-chip experiments have been helpful in estimating the hydrostatic strength $\shs$; see Section 4.2 in \cite{KLP21}. In contrast with the strength surface of most hard materials, the strength surface of some soft materials can feature a hydrostatic strength $\shs$ that is much smaller than the uniaxial and equi-biaxial tensile strengths $\sts$ and $\sbs$. That is the case, for instance, for many natural rubbers \citep{KKLP24}. The reason for this behavior is that natural rubber is nearly incompressible and hence hardly deforms and hardly stiffens when subjected to a purely hydrostatic stress. On the other hand, under uniaxial and biaxial tension, natural rubber can reach very large stretches and stiffenings before fracturing, which leads to relatively large uniaxial and equi-biaxial tensile strengths. So, especially in soft materials, the strength surfaces of different nominally elastic brittle materials can greatly vary.

In this spirit, Fig.~\ref{Fig2} presents cross sections of the strength surfaces for the titania studied by \cite{Ely72} and one of the natural rubbers (vulcanizate \texttt{D}) studied by \cite{GL59} in their famed poker-chip experiments. The former is shown in the space of principal stresses $(s_1, s_2)$ with $s_3 = 0$, while the latter is shown in the space $(s_1, s_2)$ with $s_3 = s_2$. In this figure, the solid circles correspond to the experimental data, while the solid lines have been added to help their visualization, including their stochasticity. 

Clearly, the classification of strength surfaces is a taxonomic effort, and not one that can emerge from any kind of energetic consideration.

\subsection{Nucleation from large pre-existing cracks: The Griffith energy competition}\label{Sec:Nucleation-crack}

When a specimen of a nominally elastic brittle material contains a large\footnote{``Large'' refers to large relative to the characteristic size of the underlying heterogeneities in the material under investigation. By the same token, ``small'' refers to sizes that are of the same order or just moderately larger than the sizes of the heterogeneities.} pre-existing crack, fracture may nucleate from the crack front, in other words, the crack may grow. The first systematic experimental campaign on this type of fracture nucleation is famously due to \cite{Griffith21}, who studied the nucleation of fracture from large pre-existing cracks in thin-walled cylindrical tubes and spherical bulbs made of glass that were subjected to internal pressure. For soft materials, the first experiments on this type of fracture nucleation can be traced back to the work of \cite{Busse34}, who carried out single edge notch fracture tests, as well as to the more complete work of \cite{RT53}, who carried out single edge notch, as well as the so-called ``pure-shear'' and trousers fracture tests. All of those were carried out on natural rubber. 

His pioneering experiments on glass led \cite{Griffith21} to conclusively establish that a criterion based on strength could {not} possibly explain nucleation of fracture from large pre-existing cracks. Indeed, because of the sharp geometry of the cracks, the stresses around the crack fronts greatly exceeded the strength of the glass as soon as the specimens were loaded, and yet the cracks did not grow. These experiments prompted \cite{Griffith21} to introduce the idea of energy competition between bulk deformation energy and surface fracture energy. In modern parlance, the postulate is as follows in a 2D setting: under a given load and for a preset path, a crack will have length $\ell$ in its reference state if any putative crack add-length would result in an expenditure of surface energy (assumed to be proportional to the add-length) greater than the accompanying  decrease in potential energy $\mathcal{P}$ (i.e., the total stored elastic energy  minus the work done by the external forces). Letting the add-length tend to $0$, the following condition must hold for the crack to have length $\ell$:
\begin{equation}
-\dfrac{\partial \mathcal{P}}{\partial \ell}< G_{c}.\label{Gc-1}
\end{equation}
In this expression, $G_c$ is the proportionality coefficient in the surface energy, a material property referred to as the intrinsic fracture toughness, or critical energy release rate. Note that \eqref{Gc-1} says nothing about crack propagation because it is all about crack stability. Later on, it morphed into the following statement: the crack cannot grow unless\footnote{We could not locate the original ownership of statement (\ref{Gc-0}).}
\begin{equation}
-\dfrac{\partial \mathcal{P}}{\partial \ell}=G_{c}.\label{Gc-0}
\end{equation}

Since the experiments of \cite{Griffith21} on glass and also those of \cite{RT53} on natural rubber, a multitude of experiments have been conducted. The consensus is that the Griffith criticality condition (\ref{Gc-0}) is a necessary condition for the nucleation of fracture from the front of large pre-existing cracks. A variety of routine tests have been developed for the computation of the energy release rate $-\partial \mathcal{P}/\partial \ell$, from which the critical energy release rate $G_c$ can then be directly determined; see, e.g., the handbook of \cite{Tada73} for hard materials and the work of \cite{RT53} for soft materials.

This type of nucleation, which amounts to crack propagation, is that which is most successfully addressed through a variational approach and time-indexed minimization criteria. As a consequence, it is also in that regime that variational phase-field models take flight, providing in their wake an amazingly accurate predictor for crack path. This alone vindicates the huge mathematical expenditure of the last 25 years, one that is yet to come to full completion; see, e.g., \cite{Bourdin08}.

\subsection{Nucleation under states of spatially non-uniform stress: Mediation between strength and Griffith}\label{Sec:Nucleation-transition}

The two preceding subsections have focused on fracture nucleation in nominally elastic brittle materials under two opposite limiting conditions: a spatially uniform stress field and one where the stress field is that associated with a pre-existing crack. In this subsection, we summarize the existing experimental knowledge on fracture nucleation in the in-between states. These include nucleation from smooth or sharp notches, small pre-existing cracks, and from any other subregion in the body under a non-uniform state of stress.

Experiments on specimens featuring U- and V-notches \citep{Greensmith60,Andrews63,Dunn97,Gomez05}, as well as specimens featuring small pre-existing edge cracks \citep{Thomas1970,Kimoto85,Ritchie04,Chen17} have shown that nucleation of fracture from the front of the notch or crack is the result of a mediation between strength and energy competition. The same is true for fracture nucleation that occurs in any other subregion where the stress is non-uniform, such as in indentation tests \citep{Roesler56,Mouginot85,Lawn98} and Brazilian tests \citep{Sato79,Bisai19,Sheikh19} in hard materials and in poker-chip tests \citep{GL59,Euchler20,GuoRavi23} and ``two-particle'' tests \citep{GentPark84,Poulain17,Poulain18} in soft materials.

\begin{figure}[t!]
   \centering \includegraphics[width=0.95\linewidth]{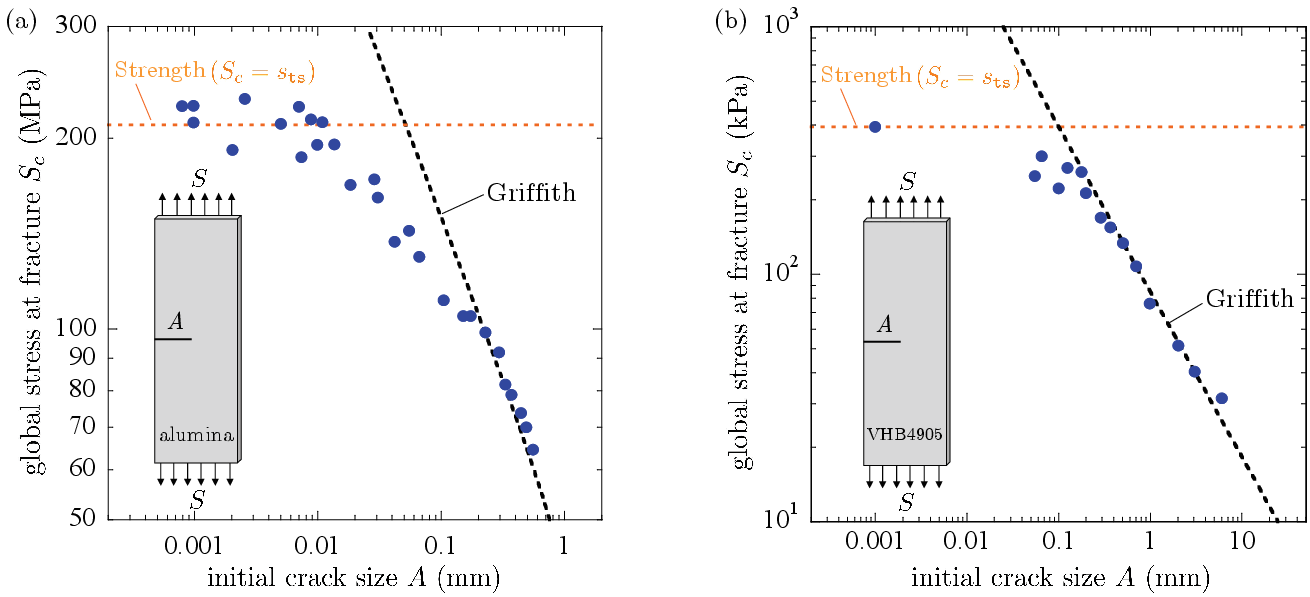}
   \caption{\small Experimental data of \cite{Kimoto85} and \cite{Chen17} for the critical global stress $S_c$ at which fracture nucleates from a pre-existing crack in single edge notch tests (schematically depicted by the inset) performed on (a) alumina, a hard ceramic, and (b) VHB 4905, a soft elastomer. The data is plotted (solid circles) as a function of the initial crack size $A$. For direct comparison, the predictions generated by the pertinent strength criterion (\ref{FPK}) and the Griffith criticality condition (\ref{Gc-0}) are included in the plots (dotted and dashed lines).}
   \label{Fig3}
\end{figure}

By way of an example, Fig.~\ref{Fig3} presents the experimental data of \cite{Kimoto85} and \cite{Chen17} for the critical global stress $S_c$ at which fracture nucleates from the pre-existing cracks in single edge notch fracture tests on an alumina ceramic and the elastomer VHB 4905. Both sets of results show that fracture nucleation for sufficiently large cracks is well described by the Griffith criticality condition (\ref{Gc-0}), while for sufficiently small cracks it is well described by the strength criterion (\ref{FPK}). In these two sets of experiments, the uniaxial tensile strength $\sts$ suffices. For crack sizes that lie between these two opposite limiting behaviors, fracture nucleation is indeed seen to occur as a mediation between the strength criterion (\ref{FPK}) and the Griffith criticality condition (\ref{Gc-0}).  

The experimental results in Fig.~\ref{Fig3} are also helpful in illustrating two additional features of fracture nucleation in elastic brittle materials under states of spatially non-uniform stress. First, the violation (\ref{FPK-bound}) of the strength of the material is a necessary condition for fracture nucleation to occur. Because of the sharpness of the notch, the stresses around the crack fronts greatly exceed the strength of the materials as soon as the specimens are loaded, and hence always prior to the nucleation of fracture, this irrespective of the initial crack size. As discussed in Subsection \ref{Sec: Strength ingredient} above, the first instance of the violation (\ref{FPK-bound}) of the strength of the material is a necessary and sufficient condition for nucleation under states of spatially uniform stress. Under states of spatially non-uniform stresses, that is no longer so and violation (\ref{FPK-bound}) is merely necessary.
     
Second, the mediation between the strength and the Griffith energy competition that determines fracture nucleation in the experiments presented in Fig.~\ref{Fig3} involves a material length scale. 
More generally, when the applied global stress is triaxial and not merely uniaxial as in Fig.~\ref{Fig3}, the scale must be tied to a stress-indexed family of material length scales. This comes about because of the different units of the strength ($force/length^2$), the elasticity ($force/length^2$), and the toughness ($force/length$), together with the fact that the strength is not a scalar quantity but an entire surface in stress space. What this family, say $\ell_{\bfS}$, should be is at present an open question. What is clear, nonetheless, is that $\ell_{\bfS}$ is not described by a single value, as often suggested in the literature, but by a range of values dependent on the strength surface $\mathcal{F}(\bfS)=0$.

\begin{figure}[t!]
   \centering \includegraphics[width=0.9\linewidth]{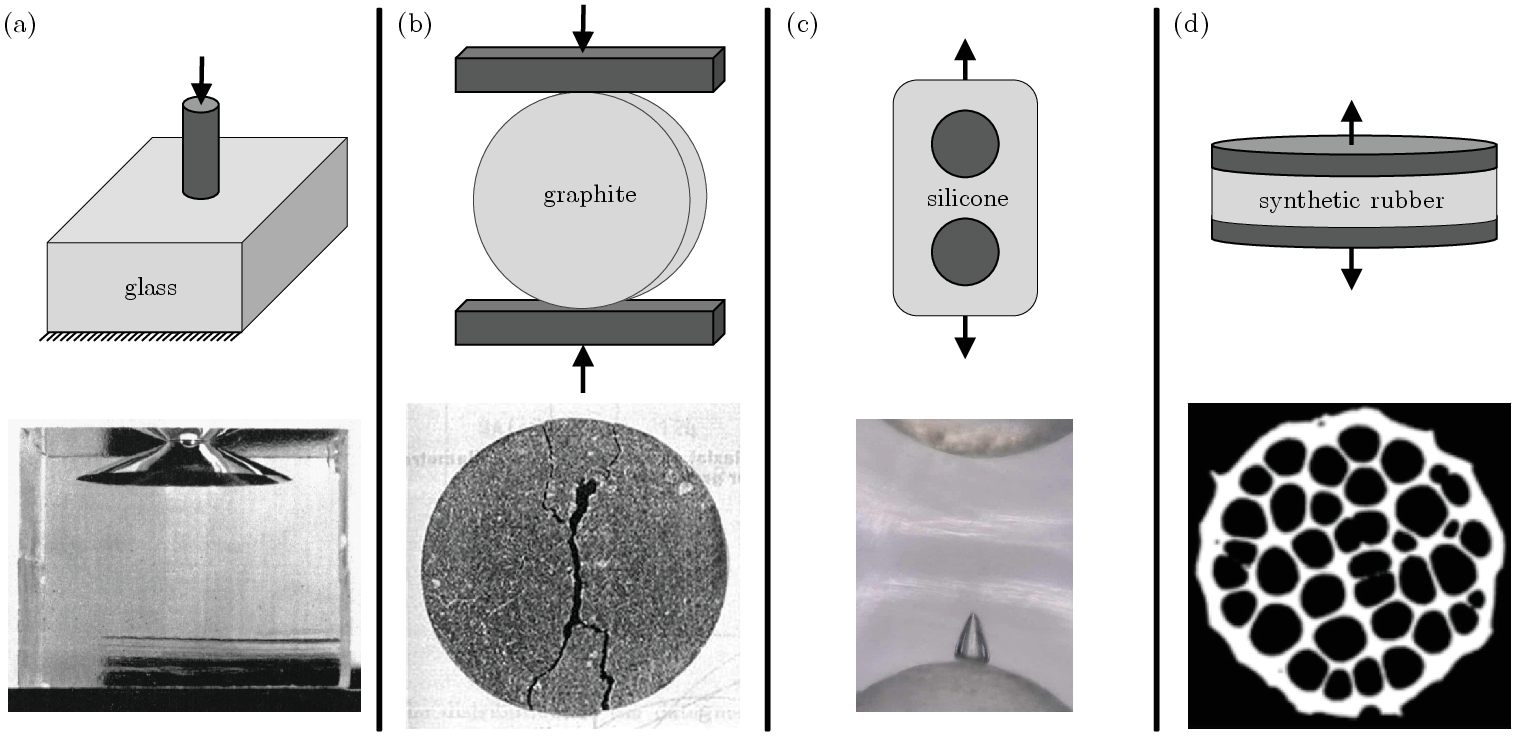}
   \caption{\small Examples of images of fracture nucleation under states of spatially non-uniform stress in popular experiments for hard and soft materials. (a) Indentation of glass with a flat-ended cylindrical indentor \citep{Roesler56}. (b) Brazilian test on graphite \citep{Sato79}.  (c) ``Two-particle'' test on a silicone elastomer \citep{Poulain17}. (d) Poker-chip test on a synthetic rubber \citep{Euchler20}. Each example includes a schematic of the test.}
   \label{Fig4}
\end{figure}

Figure \ref{Fig4} presents four other experimental results that illustrate the above for both hard and soft materials. Part (a) of Fig.~\ref{Fig4} shows the image of the crack nucleated in a block of glass that has been indented with a flat-ended cylindrical indentor \citep{Roesler56}, while part (b) shows the image of the cracks nucleated in a Brazilian test on graphite \citep{Sato79}. Moreover, part (c) shows the image in the deformed configuration of the crack nucleated in a ``two-particle'' test on a silicone elastomer \citep{Poulain17}, while part (d) shows the X-ray image, also in the deformed configuration, of the multiple cracks nucleated across the midplane of the specimen in a poker-chip test on a synthetic rubber \citep{Euchler20}. Elastic analyses of these types of tests corroborate that  the stresses in the subregions where fracture nucleates exceed the strength of the material at the time of fracture nucleation; see, e.g., \cite{KLP20}, \cite{KRLP22}, \cite{KLDLP24}, and \cite{KKLP24}.

Summing up, the experiments outlined above indicate that when a macroscopic sample of a nominally elastic brittle material is subjected to a state of spatially non-uniform stress, fracture will nucleate in subregions where the strength of the material has been exceeded, provided that said regions are sufficiently large, typically larger than a pertinent material length scale. That scale $\ell_{\bfS}$ should in turn be born out of a mediation between strength and Griffith. A mathematical formulation of sharp fracture that would encompass such a mediation while remaining  consistent with the limiting cases discussed in the two preceding subsections is wanting. 


\section{Variational phase-field models for elastic brittle materials}\label{Sec: The P-F model}

Leveraging decades of progress in the Calculus of Variations, \cite{Francfort98} introduced a mathematically consistent variational formulation of the Griffith postulate for crack growth under quasi-static loading conditions.\footnote{On page 166 of \cite{Griffith21}, the author posits that ``\emph{In an elastic solid body deformed by specified forces applied at its surface, the sum of the potential energy of the applied forces and the strain energy of the body is diminished or unaltered by the introduction of a crack whose surfaces are traction-free}''.}  For an isotropic linear elastic brittle material occupying an open bounded domain $\Omega_0\subset \mathbb{R}^3$ in its undeformed and stress-free configuration, the formulation, in its time-discretized version, is simple. Denoting by $W:M^{3\times3}\to\mathbb{R}$ the elastic energy density, or elastic energy for short, by $\Gamma(t)$ the crack surface at time $t$, by $\bfE(\bfu):=1/2(\nabla\bfu+\nabla\bfu^T)$  the symmetrized gradient of the displacement field $\bfu:\Omega_0\times[0,T]\to \mathbb{R}^3$, and by $\overline{\bfu}$ a time-dependent displacement prescribed on a part $\partial\Omega_0^\mathcal{D}$ of the boundary, and assuming further that no other loading processes are applied to $\Omega_0$, then, at times $t_k\in\{0=t_0,t_1,...,t_m,t_{m+1},...,t_M=T\}$, the pair $(\bfu_k:=\bfu_k(\bfX,t_k), \Gamma_k:=\Gamma(t_k))$ minimizes
\begin{equation}
\mathcal{E}(\bfu,\Gamma):=\int_{\Omega_0\setminus\Gamma}W(\bfE(\bfu))\,{\rm d}\bfX+G_c\mathcal{H}^{2}\left(\Gamma\right)\mbox{ among all  }(u,\Gamma)\mbox{ with }\left\{\begin{array}{l}\bfu=\overline{\bfu}(\bfX,t_k)\,{\rm on \;}\,\partial\Omega_0^\mathcal{D}\setminus\Gamma\\[2mm] \Gamma\supset\Gamma_{k-1}\end{array}\right. .\label{FM98}
\end{equation}
In this expression, $\mathcal{H}^{2}(\Gamma)$ stands for the $2$--dimensional Hausdorff measure (the surface measure) of the unknown crack $\Gamma$. The absence of force loads is not coincidental but an unfortunate necessity because their presence would drive the minimum of (\ref{FM98}) to $-\infty$, thereby rendering the whole process meaningless; for more on this, including an alternative formulation, see \cite{Larsen21}. 

Formulation (\ref{FM98}) is not  amenable to numerical implementation, nor are any approximate formulations, whenever approximation is in the sense of $\Gamma$-convergence.\footnote{$\Gamma$-convergence of $\mathcal{E}^\e$ to $\mathcal{E}$ essentially means that global minimizers of $\mathcal{E}^\e$ converge to global minimizers of $\mathcal{E}$, and all global minimizers of $\mathcal{E}$ are limits of global minimizers of $\mathcal{E}^\e$. Outside of global minimality, $\Gamma$-convergence is of little relevance.} However, as an alternative to global minimization, phase-field approximations are naturally implementable based on separate minimization (i.e., Nash equilibrium), which we describe below. This separate minimization, while it removes the link to the sharp theory, yields results that are consistent with experimental observations. In the sequel, we provide a review of two classes of phase-field approximations: the variational phase-field models without and those with energy splits.

\subsection{The original variational phase-field models}

The first class of variational phase-field models that were introduced to approximate (\ref{FM98}) are of the form \citep{Bourdin00,Bourdin08}
\begin{align}
&(\bfu^\e_k,v^\e_k)=\underset{\begin{subarray}{c}
  \bfu=\overline{\bfu}(\bfX,t_k)\,{\rm on}\,\partial\Omega_0^\mathcal{D} \\[1mm]
  0 \leq v\leq v_{k-1}\leq 1
  \end{subarray}}{\arg\min}\,\mathcal{E}^{\e}(\bfu,v):=\displaystyle
  \int_{\Omega_0}g(v)W(\bfE(\bfu))\,{\rm d}\bfX-\displaystyle
  \int_{\partial\Omega^{\mathcal{N}}_0}\overline{\textbf{s}}(\bfX,t_k)\cdot\bfu\,{\rm d}\bfX+\nonumber\\
  &\hspace{5.75cm}\dfrac{G_c}{4 c_s}\int_{\Omega_0}\left(\dfrac{s(v)}{\e}+\e\nabla v\cdot\nabla v\right)\,{\rm d}\bfX,\label{BFM00}
\end{align}
where $\e>0$ is a regularization length, $v$ is an order parameter or phase-field variable taking values in $[0,1]$, $g$ and $s$ are continuous strictly monotonic functions such that $g(0)=0$, $g(1)=1$, $s(0)=1$, $s(1)=0$, and $c_s:=\int_0^1\sqrt{s(z)}{\rm d}z$ is a normalization parameter \citep{Braides98}. Note that we have now included a boundary force load $\overline{\textbf{s}}(\bfX,t)$ on $\partial\Omega_0^{\mathcal{N}}=\partial\Omega_0\setminus\partial{\Omega}_0^{\mathcal{D}}$ in the variational problem (\ref{BFM00}), since these loads can be handled by the separate minimization process described next. 

Observe that:

\begin{enumerate}[label=\itshape\roman*.]

\item{The energy functional $\mathcal{E}^{\e}(\bfu,v)$ in (\ref{BFM00}) is not convex so that finding a minimizer is an impossible task at present. However, it is separately convex and thus lends itself to staggered minimization. Specifically,  knowing $(\bfu^\e_{k-1},v^\e_{k-1})$ and setting $(\bfu_0, \bfv_0):=(\bfu^\e_{k-1},v^\e_{k-1})$, one minimizes successively $\mathcal{E}^{\varepsilon}(\cdot\;, v_{i-1})$ with $\bfu=\overline{\bfu}(\bfX,t_k)\,{\rm on}\,\partial\Omega_0^\mathcal{D}$, thereby producing $\bfu_i$ and $\mathcal{E}^{\varepsilon}(\bfu_{i}, \;\cdot)$ with $v\le v^\e_{k-1}$, thereby producing $\bfv_i$  until convergence. The result can be shown to be a critical point for $\mathcal{E}^{\e}(\cdot,\cdot)$ under the constraints  $\bfu=\overline{\bfu}(\bfX,t_k)\,{\rm on}\,\partial\Omega_0^\mathcal{D}$, $v\le v^\e_{k-1}$, but nothing more.

In other words, even in the absence of boundary force loads when $\overline{\textbf{s}}(\bfX,t)=\textbf{0}$, the solutions generated from the class of variational phase-field models (\ref{BFM00}) by alternating minimization need {not} be global minimizers and hence need {not} be approximations of the solutions of the variational theory of \cite{Francfort98}, or even separate minimizers; }

\item{The two most common choices for the degradation and surface regularization functions $g(v)$ and $s(v)$, and hence the constant $c_s$, in the class of variational phase-field models (\ref{BFM00}) are
\begin{align}\label{AT1AT2}
\texttt{AT}_{1}\,\left\{
\begin{array}{l}
g(v)=v^2\vspace{0.2cm}\\
s(v)=1-v
\vspace{0.2cm}\\
c_s=\dfrac{2}{3}\end{array}\right.\qquad {\rm and}\qquad 
\texttt{AT}_{2}\,\left\{
\begin{array}{l}
g(v)=v^2\vspace{0.2cm}\\
s(v)=(1-v)^2
\vspace{0.2cm}\\
c_s=\dfrac{1}{2}\end{array}\right. .
\end{align}
They are referred to as the $\texttt{AT}_{1}$ and $\texttt{AT}_{2}$ variational phase-field models in reference to the work of \cite{AT90,AT92}. The $\texttt{AT}_{1}$ variational phase-field model is generally preferred over $\texttt{AT}_{2}$; and}

\item{Evidently, since conceived as approximations of the variational formulation (\ref{FM98}) of \cite{Francfort98}, the class of variational phase-field models (\ref{BFM00}) do not account for the strength surface $\mathcal{F}(\bfS)=0$ of the material.}
    
\end{enumerate}
    
As already pointed out in the Introduction, the presence of a small but finite regularization length $\varepsilon$ in (\ref{BFM00}) opens a window into a type of fracture nucleation that does not exist in the sharp formulation (\ref{FM98}). This feature, brought to the fore by numerical experiments,  prompted several works in which $\varepsilon$ is viewed not as a mere regularization length, but rather as a material length scale \citep{Freddi10,Marigo11}. The hope was that the variational phase-field models (\ref{BFM00}), with $\varepsilon$ fixed to some material-specific value, would describe not only fracture propagation, but also fracture nucleation.
    
Following that guiding principle, \cite{Tanne18} proposed the prescription
\begin{align}\label{eps-AT1}
\varepsilon=\dfrac{3 G_c E}{8\sts^2}
\end{align}
for use in the $\texttt{AT}_{1}$ variational phase-field model. Here is how this comes about. Recognizing that the uniaxial tensile strength $\sts$ is a material property, one considers the problem of a bar under uniaxial tension. The phase field solution $(\bfu^\e(t), v^\e(t))$ (assumed to be uniquely determined) is computed and one looks at the instant $t^\e_c$ at which a spatially uniform $v^\e(t)$ loses linear stability. One then computes the corresponding stress $S^\e_c$ and equates it to $\sts$. This will produce a set value for $\e$, namely, \eqref{eps-AT1}.

Comparing the predictions of the resulting model with experimental results on fracture nucleation from U- and V-notches for a handful of hard materials under tension, an apparent agreement of the predictions with those experiments led \cite{Tanne18} to state the following in their Abstract:

\medskip

\noindent \emph{``Our main claim, supported by validation and verification in a broad range of materials and geometries, is that crack nucleation can be accurately predicted by minimization of a
nonlinear energy in variational phase field models, and does not require the introduction of ad-hoc criteria.''}

\medskip

\noindent This claim is in contradiction with the experimental observations reviewed above, which show that the strength of nominally elastic brittle materials is one of the macroscopic material properties that govern fracture nucleation and that such material property is characterized by an entire strength surface $\mathcal{F}(\bfS)=0$ in stress space that is independent from elasticity and toughness.

In Section \ref{Sec: The evidence} below, we confront the $\texttt{AT}_{1}$ variational phase-field with a wider range of experimental observations and show that the apparent agreement with experiments found by \cite{Tanne18} was only so because of the very specific type and limited set of experiments that were considered in that work.

\subsection{Variational phase-field models with energy splits}

In the very first numerical experiments using the $\texttt{AT}_{2}$ variational phase-field model (see Fig.~4(f) in \cite{Bourdin00}), it was shown that cracks could grow from the front of large pre-existing cracks into regions with compressive strain fields. Cracks in actual materials are not expected to grow there. \cite{Bourdin00} argued that the reason for this unphysical behavior  is the lack of material impenetrability in formulation (\ref{BFM00}). While not accounting for material impenetrability is definitely an issue that needs to be addressed, the root cause for such an unphysical behavior is elsewhere: the elastic energy $W(\bfE)$ entering the competition in the class of phase-field models (\ref{BFM00}) cannot distinguish between tensile and compressive strains, even when material impenetrability is not violated. 

As an indirect approach to material impenetrability,\footnote{Material impenetrability is first and foremost a kinematical constraint and not an energetic one; see, e.g., \cite{GP08}.}  \cite{Amor09} proposed to modify the class of phase-field models (\ref{BFM00}) through a split of the elastic energy $W(\bfE)=W^+(\bfE)+W^-(\bfE)$ into a ``tensile'' or ``degradable'' part $W^+(\bfE)$ and a ``compressive'' or ``residual'' part $W^-(\bfE)$, only allowing the ``tensile'' part to be degraded by $g(v)$. Around the same time, \cite{Miehe10} proposed a different energy split that was not motivated by material impenetrability but that would target the unphysical growth of cracks in regions of large compressive strains. Thereafter, many different energy splits have been proposed in the literature; see, e.g., the reviews included in \cite{Wick22} and \cite{Perego24}. They all read as
\begin{align}
&(\bfu^\e_k,v^\e_k)=\underset{\begin{subarray}{c}
  \bfu=\overline{\bfu}(\bfX,t_k)\,{\rm on}\,\partial\Omega_0^\mathcal{D} \\[1mm]
  0 \leq v\leq v_{k-1}\leq 1
  \end{subarray}}{\arg\min}\,\mathcal{E}_{\texttt{Split}}^{\e}(\bfu,v):=\displaystyle
  \int_{\Omega_0}\left(g(v)W^+(\bfE(\bfu))+W^-(\bfE(\bfu))\right)\,{\rm d}\bfX-\displaystyle
  \int_{\partial\Omega^{\mathcal{N}}_0}\overline{\textbf{s}}(\bfX,t_k)\cdot\bfu\,{\rm d}\bfX+\nonumber\\
  &\hspace{6.25cm}\dfrac{G_c}{4 c_s}\int_{\Omega_0}\left(\dfrac{s(v)}{\e}+\e\nabla v\cdot\nabla v\right)\,{\rm d}\bfX,\label{W-Split}
\end{align}
where the ``tensile'' part $W^+(\bfE)$ and the ``compressive'' part $W^-(\bfE)$ are such that  $W^-(\bfE)\geq 0$ and $W^+(\bfE)+W^-(\bfE)=W(\bfE)$, but are otherwise arbitrary.

Note that:

\begin{enumerate}[label=\itshape\roman*.]

\item{Variational convergence of split phase-field models to any kind of sharp fracture model is lacking except in 2D and for a very specific class of split \citep{CCF18};}
  
\item{Much like $\mathcal{E}^{\e}(\bfu,v)$ in (\ref{BFM00}), the energy functional $\mathcal{E}_{\texttt{Split}}^{\e}(\bfu,v)$ in (\ref{W-Split}) is not convex. It is at best separately convex and thus amenable to staggered minimization; and}
    
\item{Constructed as modifications of the variational phase-field models (\ref{BFM00}), the class of variational phase-field models (\ref{W-Split}) incorporates the elasticity and toughness of the material, but not its strength surface $\mathcal{F}(\bfS)=0$.}
    
\end{enumerate}
    
As already recalled in the Introduction and as further elaborated in the next section, \emph{no} variational phase-field model, without or with energy splits, can possibly describe fracture nucleation in general. Yet, variational phase-field models continue to be insisted upon for modeling nucleation. In that vein, a notable recent contribution by \cite{Maurini24}, labeled the \texttt{star-convex} variational phase-field model, stands out because, according to the authors (see the Introduction in \cite{Maurini24}): 

\medskip

\noindent \emph{``...this model is still based on energy decomposition, but it is specifically designed to satisfy the desired requirements for both nucleation and propagation.''}

\medskip

\noindent In the \texttt{star-convex} variational phase-field model, the ``tensile'' and ``compressive'' parts $W^+(\bfE)$ and $W^-(\bfE)$ are given by

\begin{align}\label{W-Star-Convex}
\left\{\hspace{-0.1cm}\begin{array}{l}
W^+(\bfE)=\mu\left(\,{\rm tr}\, \bfE^2-\dfrac{1}{3}({\rm tr}\, \bfE)^2\right)+\dfrac{\kappa}{2}\left(({\rm tr}\, \bfE^+)^2-\gamma^\star({\rm tr}\, \bfE^-)^2\right)\vspace{0.25cm}\\
W^-(\bfE)=(1+\gamma^\star)\dfrac{\kappa}{2}({\rm tr}\, \bfE^-)^2\end{array}\right.\, {\rm with}\quad \left\{\hspace{-0.15cm}\begin{array}{l}{\rm tr}\,\bfE^+:=\left\{\hspace{-0.1cm}\begin{array}{ll}
{\rm tr}\,\bfE, &{\rm tr}\,\bfE\geq0\vspace{0.2cm}\\
0, &{\rm else}
\end{array}\right.\vspace{0.2cm}\\{\rm tr}\,\bfE^-:=\left\{\hspace{-0.1cm}\begin{array}{ll}
{\rm tr}\,\bfE, &{\rm tr}\,\bfE<0\vspace{0.2cm}\\
0, &{\rm else}\end{array}\right.\end{array}\right.,
\end{align}
where the parameter $\gamma^\star$ satisfies $\gamma^\star\geq -1$, and where $\mu$ and $\kappa$ denote the shear and bulk moduli of the material.\footnote{Recall that $\mu$ and $\kappa$ are given in terms of the Young's modulus and the Poisson's ratio by $\mu=E/(2(1+\nu))$ and $\kappa=E/(3(1-2\nu))$.}

In the next section, we confront the \texttt{star-convex} variational phase-field model (\ref{W-Split}) with (\ref{W-Star-Convex}) directly with experimental observations and show that, as expected, fracture nucleation is beyond its reach.

\section{Variational phase-field models vs. experimental observations}\label{Sec: The evidence}

Per the experimental observations summarized in Section \ref{Sec: Experiments}, three necessary ingredients must be accounted for by any phase-field model, be it variational or not, if it is  to potentially describe fracture nucleation in elastic brittle materials. These are:
\begin{itemize}

\item{Accounting for the elastic energy $W(\bfE)$, the strength surface $\mathcal{F}(\bfS)=0$, and the toughness $G_c$ of the material;}

\item{Localization of the phase field $v$ whenever a macroscopic piece of the material is subjected to any uniform stress $\bfS$ that exceeds the strength surface $\mathcal{F}(\bfS)=0$ of the material; and}

\item{Having the Griffith energy competition as a descriptor of nucleation from a large pre-existing crack.}

\end{itemize}
Failure to satisfy any of these requirements would prevent the model from describing fracture nucleation even in the simplest of scenarios, that is, under a spatially uniform stress and/or from large pre-existing cracks.

The choices of degradation and surface regularization functions $g(v)$ and $s(v)$ in (\ref{BFM00}) and (\ref{W-Split}), as well as the choice of the split $W^+(\bfE)$ and $W^-(\bfE)$ of the elastic energy in (\ref{W-Split}), can result in variational phase-field models for which the phase field $v$ does not properly localize and/or for which nucleation from a large pre-existing crack is not described by the Griffith energy competition.\footnote{Localization of the phase field $v$ is essential. As illustrated, for example, in Fig.~\ref{Fig1}(a), nominally elastic brittle materials deform and crack. Given that phase-field models describe cracks via a phase field $v$ that localizes in space near $v=0$, a phase field $v<1$ that does not localize is {not} descriptive of reality.}$^{,}$\footnote{Other types of phase-field models, like those where the elastic energy is replaced by a non-energetic driving force in the evolution equation for the phase field do not typically lead to the proper localization of the phase field and/or do not typically describe nucleation from a large pre-existing crack according to the Griffith energy competition; see, for example, \cite{Miehe15,Weinberg19}.}

However, even a proper choice of $g(v)$ and $s(v)$ in (\ref{BFM00}) and of $g(v)$, $s(v)$, $W^+(\bfE)$, and $W^-(\bfE)$ in (\ref{W-Split}) will result in a variational phase-field model that fails to satisfy the first of the three necessary ingredients. This is because the variational phase-field models  (\ref{BFM00}) and (\ref{W-Split}) cannot account for the strength surface $\mathcal{F}(\bfS)=0$ as an independent material property. This is the fundamental reason why variational phase-field models cannot possibly describe fracture nucleation.

As first illustrated by \cite{KBFLP20} and as recalled in the Introduction, the variational phase-field models  (\ref{BFM00}) and (\ref{W-Split}) can be made to account for certain types of strength surfaces by appropriately choosing the value of the regularization length $\varepsilon$ and the type of energy split. However, the strength surfaces that are generated from such an 
approach are extremely limited in their functional form. What is more, they are non-physical in certain regions of stress space. In short, they are not representative of actual materials. This is because those surfaces can only be written in terms of the elastic energy $W(\bfE)$ and the toughness $G_c$ of the material, instead of being truly independent material properties. 

Next, by way of examples for both different classes of variational phase-field models (\ref{BFM00}) and (\ref{W-Split}), we present the strength surfaces generated by the \texttt{AT}$_1$ variational phase-field model (\ref{BFM00}) with (\ref{AT1AT2})$_1$ and by the \texttt{AT}$_1$ version of the \texttt{star-convex} variational phase-field model (\ref{W-Split}) with (\ref{AT1AT2})$_1$ and (\ref{W-Star-Convex}). After discussing their main features, we confront the resulting strength surfaces with experimental data for four materials of common use.

\subsection{The strength surface generated by the \emph{\texttt{AT}$_1$} variational phase-field model}

The strength surfaces generated by variational phase-field models without energy splits (\ref{BFM00}) --- assuming a correct localization of $v$ --- are simply defined by the Euler-Lagrange equation of the minimization problem (\ref{BFM00}) associated with variations in the phase field $v$ when evaluated at $v=1$ and at spatially uniform strains $\bfE$, and hence at spatially uniform stresses $\bfS$.\footnote{As discussed in Appendix A in \cite{KBFLP20}, the loss of linear stability of the spatially affine/homogeneous solution pair ($\bfu,v$) is a necessary condition for localization and that loss eventually leads to this statement.} 

For the case of the \texttt{AT}$_1$ variational phase-field model, the Euler-Lagrange equation associated with variations in $v$, evaluated at  $v=1$ and at spatially uniform strains $\bfE$, reads simply
\begin{equation*}
2 W(\bfE)-\dfrac{3G_c}{8\varepsilon}=0.
\end{equation*}
Rewriting the constant strain $\bfE$ in terms of the associated constant stress $\bfS=E/(1+\nu)\bfE+\nu E/((1+\nu)(1-2\nu))({\rm tr}\bfE)\bfI$ yields as strength surface
\begin{equation}\label{F-AT1}
\mathcal{F}^{{\texttt{AT}}_{1}}(\bfS)=\dfrac{\mathcal{J}_2}{\mu}+\dfrac{\mathcal{I}_1^2}{9\kappa}-\dfrac{3 G_c}{8\varepsilon}=0,
\end{equation}
where we recall that the stress invariants $\mathcal{I}_1$ and $\mathcal{J}_2$ are given by (\ref{Inv-S}). 

The surface $\mathcal{F}^{{\texttt{AT}}_{1}}(\bfS)=0$ is not an independent material property, but one that is subordinate to the elasticity and the toughness of the material, even if $\varepsilon$ is viewed as a free parameter. Note also that the strength surface $\mathcal{F}^{{\texttt{AT}}_{1}}(\bfS)=0$ exhibits a very specific functional dependence on the stress invariants $\mathcal{I}_1$ and $\mathcal{J}_2$. Such a dependence is inconsistent with experimental observations as obviated in the special case of hydrostatic tensile loading, when $\bfS={\rm diag}(s>0,s>0,s>0)$, for which (\ref{F-AT1}) predicts
\begin{equation*}
	\shs=\sqrt{\dfrac{3 G_c \kappa}{8\varepsilon}}
\end{equation*}
for the hydrostatic strength $\shs$ of the material. Thus, according to the \texttt{AT}$_1$ variational phase-field model, the hydrostatic strength $\shs$ increases with increasing values of the bulk modulus $\kappa$ of the material. In particular, incompressible materials ($\kappa=+\infty$) are predicted to be infinitely strong ($\shs=+\infty$) under hydrostatic tension. This is nonsensical.

When considered as a material length scale, $\varepsilon$ can be viewed as the  sole tunable parameter in (\ref{F-AT1}), one whose value can be selected so that the strength surface $\mathcal{F}^{{\texttt{AT}}_{1}}(\bfS)=0$ matches the actual strength surface $\mathcal{F}(\bfS)=0$ of the material of interest at a single point of choice in stress space. For example, the prescription (\ref{eps-AT1}) forces $\mathcal{F}^{{\texttt{AT}}_{1}}(\bfS)=0$  to match the actual uniaxial tensile strength $\sts$ of the material.

\subsection{The strength surface generated by the \emph{\texttt{star-convex}} variational phase-field model}

Exactly as for the variational phase-field models without energy splits (\ref{BFM00}), the strength surfaces generated by variational phase-field models with energy splits (\ref{W-Split}),  assuming, once more, proper localization of the phase field $v$, are defined by the Euler-Lagrange equation of the minimization problem (\ref{W-Split}) associated with variations in the phase field $v$ when evaluated at $v=1$ and at spatially uniform strains $\bfE$. Upon following the argument outlined in the previous subsection, we find the following strength surface for the \texttt{star-convex} variational phase-field model:
\begin{equation}\label{F-Star}
\mathcal{F}^{\star}(\bfS)=\left\{\begin{array}{ll} \dfrac{\mathcal{J}_2}{\mu}+\dfrac{\mathcal{I}_1^2}{9\kappa}-\dfrac{3 G_c}{8\varepsilon}=0, & \mathcal{I}_1\geq0\vspace{0.2cm}\\\dfrac{\mathcal{J}_2}{\mu}-\gamma^\star\dfrac{\mathcal{I}_1^2}{9\kappa}-\dfrac{3 G_c}{8\varepsilon}=0, & \mathcal{I}_1<0 \end{array}\right.,
\end{equation}
where, once again, $\gamma^\star\geq -1$.\footnote{In all fairness, one should re-examine the analysis in Appendix A in \cite{KBFLP20} for the case where $\mathcal{I}_1<0$, which we have not done.}   

The strength surface (\ref{F-Star}) reduces identically to the strength surface (\ref{F-AT1}) generated by the \texttt{AT}$_1$ variational phase-field model when $\mathcal{I}_1=s_1+s_2+s_3\geq 0$, that is, when the volumetric part of the stress is non-negative. This implies that both strength surfaces generate the same uniaxial tensile, equi-biaxial tensile, hydrostatic tensile, and shear strengths $\sts, \sbs,\shs$, and $\sss$. By the same token, the criticisms in the previous subsection equally apply to (\ref{F-Star}).

For stress states with $\mathcal{I}_1< 0$, the strength surface (\ref{F-Star}) improves on (\ref{F-AT1}) in that it contains one additional parameter, $\gamma^\star$, whose value can be selected to match the actual strength surface $\mathcal{F}(\bfS)=0$ of the material of interest at one additional point of choice in stress space.  Since the value of $\gamma^\star$ is only relevant for points in stress space where $\mathcal{I}_1<0$, a natural point to match is that of uniaxial compressive strength $\scs$. That choice leads to the prescription 
\begin{equation*}
\gamma^\star=\left(\dfrac{3}{\mu}-\dfrac{27 G_c}{8\varepsilon\, \scs^2}\right)\kappa.
\end{equation*}
If using in turn (\ref{eps-AT1}) for $\varepsilon$ to match the uniaxial tensile strength $\sts$, this last result specializes to
\begin{equation}\label{gamma-scs}
\gamma^\star=\dfrac{1}{1-2\nu}\left(2(1+\nu)-\dfrac{3\sts^2}{\scs^2}\right).
\end{equation}
Since $\gamma^\star\geq -1$, the minimal value of uniaxial compressive strength that can be matched is $\scs=\sts$. It is also interesting to note that while the prescription (\ref{eps-AT1}) subordinates $\varepsilon$ to $E$ and $G_c$, the prescription (\ref{gamma-scs}) subordinates $\gamma^\star$ to $\nu$.

\subsection{Comparisons with experiments for four materials of common use}

\begin{figure}[t!]
   \centering \includegraphics[width=0.9\linewidth]{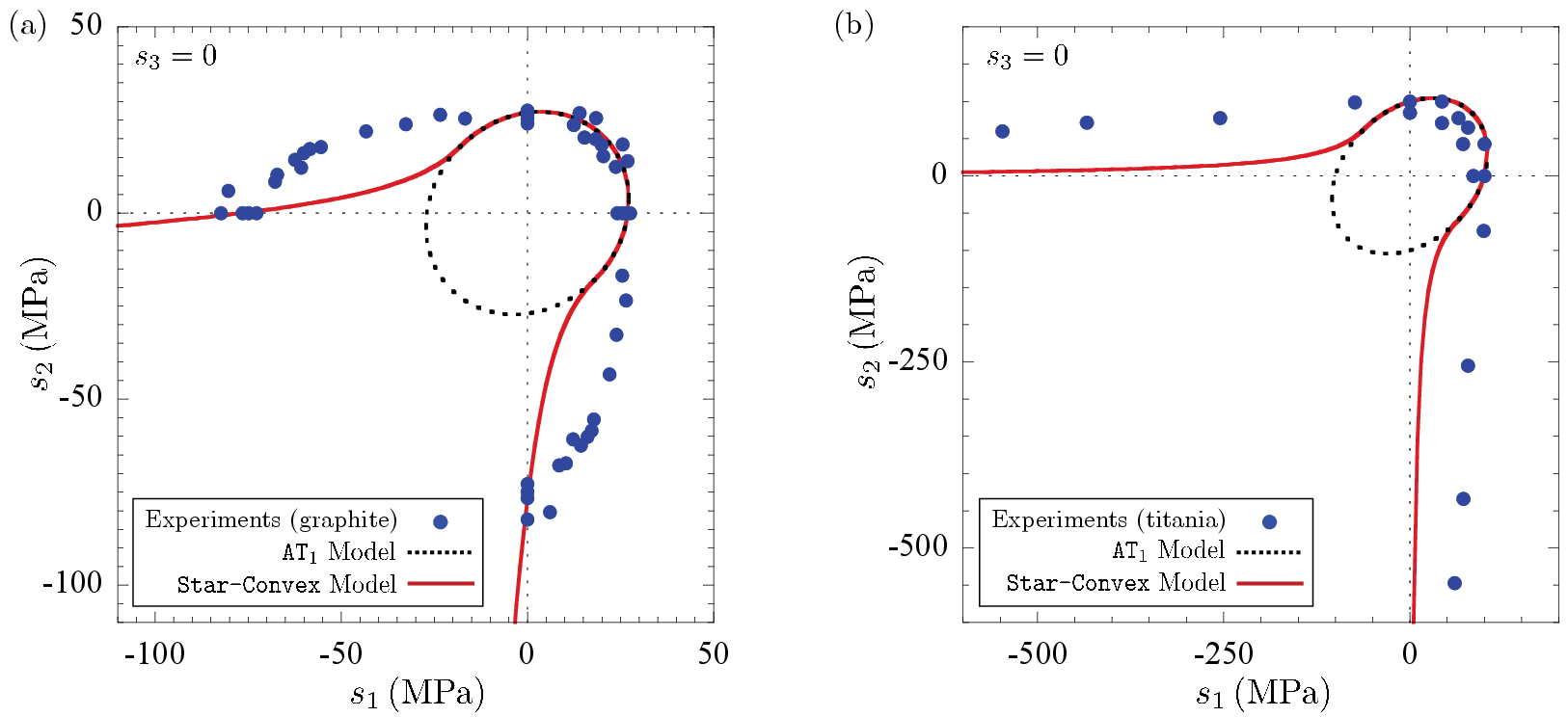}
   \caption{\small Comparisons of the strength surfaces (\ref{F-AT1}) and (\ref{F-Star}) generated by the \texttt{AT}$_1$ and \texttt{star-convex} variational phase-field models with experimental data for (a) graphite and (b) titania. The data is plotted in the space of principal stresses $(s_1, s_2)$ with $s_3 = 0$.}
   \label{Fig5}
\end{figure}

Figures \ref{Fig5} confronts the strength surfaces (\ref{F-AT1}) and (\ref{F-Star}) generated by the \texttt{AT}$_1$ and \texttt{star-convex} variational phase-field models with the experimental data for two ceramics of common use, graphite and titania, as reported by \cite{Sato87} and \cite{Ely72}. The data is plotted in the space of principal stresses $(s_1, s_2)$ with $s_3 = 0$. The relevant material constants for these two ceramics are listed in Table \ref{Table1}, which also lists the values of the parameters $\varepsilon$ and $\gamma^\star$ in the strength surfaces (\ref{F-AT1}) and (\ref{F-Star}), as obtained from (\ref{eps-AT1}) and (\ref{gamma-scs}).

\begin{table}[H]\centering
\caption{Material constants for the graphite, titania, natural rubber, and SBR considered in this work and the corresponding values for the parameters $\varepsilon$ and $\gamma^\star$ in the strength surfaces (\ref{F-AT1}) and (\ref{F-Star}) generated by the \texttt{AT}$_1$ and \texttt{star-convex} variational phase-field models.}
\begin{tabular}{r|cccc}
\hline
  & Graphite & Titania & Natural Rubber & SBR \\
\hline
Young's modulus  $E$ & $9.8$ GPa & $250$ GPa & $1.77$ MPa & $2.99$ MPa\\ \hline
Poisson's ratio $\nu$ & $0.13$ & $0.29$ & $0.4999$ & $0.4998$\\ \hline
shear modulus  $\mu$ & $4.3$ GPa & $97$ GPa & $0.588$ MPa & $1$ MPa\\ \hline
bulk modulus $\kappa$ & $4.4$ GPa & $198$ GPa & $2.2$ GPa & $2.2$ GPa\\ \hline
uniaxial tensile strength $\sts$ & $27$ MPa & $100$ MPa & $9$ MPa & $26.6$ MPa\\ \hline
uniaxial compressive strength $\scs$ & $77$ MPa & $1232$ MPa & $1000$ MPa & $1000$ MPa\\ \hline
toughness $G_c$& $91$ N/m & $36$ N/m & $100$ N/m & $100$ N/m\\ \hline
$\varepsilon$ & $0.46$ mm & $0.34$ mm & $0.82$ $\mu$m & $0.16$ $\mu$m\\ \hline
$\gamma^\star$ & $2.56$  & $6.10$  & $11217$  & $6626$ \\ \hline
\end{tabular} \label{Table1}
\end{table}

Figure \ref{Fig5} calls for the following observations. First, qualitatively, the shapes of the strength surfaces (\ref{F-AT1}) and (\ref{F-Star}) are different from those of the experimental data, even in the admittedly limited subspace of plane stresses ($s_3=0$) explored by the experiments. 

Then, quantitatively, the agreement with the non-fitted experimental data varies by quadrant. In the first quadrant $(s_1\geq 0,s_2\geq 0)$, the agreement happens to be good, more so for graphite than for titania. In the third quadrant $(s_1\leq 0,s_2\leq 0)$, experimental data are not available. However, conventional wisdom dictates that these ceramics should be very strong under biaxial compression, which is not the case for the strength surface (\ref{F-AT1}). A similar disagreement may be true for the strength surface (\ref{F-Star}), but the lack of experimental data does not allow us to reach a conclusion. Finally, in the second quadrant $(s_1\leq 0,s_2\geq 0)$ --- which, from isotropy, is the same as the fourth quadrant $(s_1\geq 0,s_2\leq 0)$ --- both strength surfaces (\ref{F-AT1}) and (\ref{F-Star}) disagree with the experimental data, more so (\ref{F-AT1}) than (\ref{F-Star}).  

\begin{table}[b!]\centering
\caption{Difference between the strength surfaces (\ref{F-AT1}) and (\ref{F-Star}) and the experimental data for graphite, titania, natural rubber, and SBR in terms of the measures (\ref{R-measures}).}
\begin{tabular}{r|cccc}
\hline
  & Graphite & Titania & Natural Rubber & SBR \\
\hline
$\mathcal{R}^{\texttt{AT}_1}_{max}$ & $2.96$ & $12.32$ & $110$ & $1.88$\vspace{0.1cm}\\ \hline
$\mathcal{R}^\star_{max}$ & $1.55$ & $4.31$ & $110$ & $1.88$\vspace{0.1cm}\\ \hline
\end{tabular} \label{Table2}
\end{table}
So as to quantify the actual difference between the strength surfaces (\ref{F-AT1}) and (\ref{F-Star}) and the experimental data, we consider the maximum radial ratios 
\begin{equation}\label{R-measures}
\mathcal{R}^{\texttt{AT}_1}_{max}=\max\left\{\max_{\theta,\,\varphi}\dfrac{\mathcal{R}^{data}(\theta,\varphi)}{\mathcal{R}^{\texttt{AT}_1}(\theta,\varphi)},
\max_{\theta,\,\varphi}\dfrac{\mathcal{R}^{\texttt{AT}_1}(\theta,\varphi)}{\mathcal{R}^{data}(\theta,\varphi)}\right\}\; {\rm and}\; \mathcal{R}^\star_{max}=\max\left\{\max_{\theta,\,\varphi}\dfrac{\mathcal{R}^{data}(\theta,\varphi)}{\mathcal{R}^\star(\theta,\varphi)},
\max_{\theta,\,\varphi}\dfrac{\mathcal{R}^\star(\theta,\varphi)}{\mathcal{R}^{data}(\theta,\varphi)}\right\}
\end{equation}
in the space of principal stresses as measures of the difference. Here,
\begin{equation*}
\left\{\begin{array}{ll}\mathcal{R}^{data}(\theta,\varphi)=\sqrt{s_1^2+s_2^2+s_3^2}, & (s_1,s_2,s_3)\in\mathcal{S}^{data}\vspace{0.2cm}\\\mathcal{R}^{\texttt{AT}_1}(\theta,\varphi)=\sqrt{s_1^2+s_2^2+s_3^2}, & (s_1,s_2,s_3)\in\mathcal{S}^{\texttt{AT}_1}\vspace{0.2cm}\\
\mathcal{R}^\star(\theta,\varphi)=\sqrt{s_1^2+s_2^2+s_3^2}, & (s_1,s_2,s_3)\in\mathcal{S}^\star\end{array}\right.,
\end{equation*}
stand for the radial distances from the origin $(s_1,s_2,s_3)=(0,0,0)$ to points in the surfaces $\mathcal{S}^{data}=\left\{(s_1,s_2,s_3):\right.$ $\left.\mathcal{F}^{data}({\rm diag}(s_1,s_2,s_3))=0\right\}$, $\mathcal{S}^{\texttt{AT}_1}=\left\{(s_1,s_2,s_3): \mathcal{F}^{\texttt{AT}_1}({\rm diag}(s_1,s_2,s_3))=0\right\}$, and $\mathcal{S}^\star=\left\{(s_1,s_2,\right.$ $\left.s_3): \mathcal{F}^\star({\rm diag}(s_1,s_2,s_3))=0\right\}$ along the directions $\theta=\tan^{-1}(s_2/s_1)$ and $\varphi=\cos^{-1}(s_3/\sqrt{s_1^2+s_2^2+s_3^2})$ in the space of principal stresses $(s_1,s_2,s_3)$. Table \ref{Table2} presents the values of the maximum radial ratios. For graphite, there is a direction in which the strength surface (\ref{F-AT1}) is 2.96 times smaller than the corresponding experimental data, while in a different direction the strength (\ref{F-Star}) is 1.55 times smaller than the corresponding experimental data. The comparison for titania is even more striking, as there are directions in which the strength surfaces (\ref{F-AT1}) and (\ref{F-Star}) are 12.32 and 4.31 times smaller than the experimental data, respectively.

\begin{figure}[t!]
   \centering \includegraphics[width=0.9\linewidth]{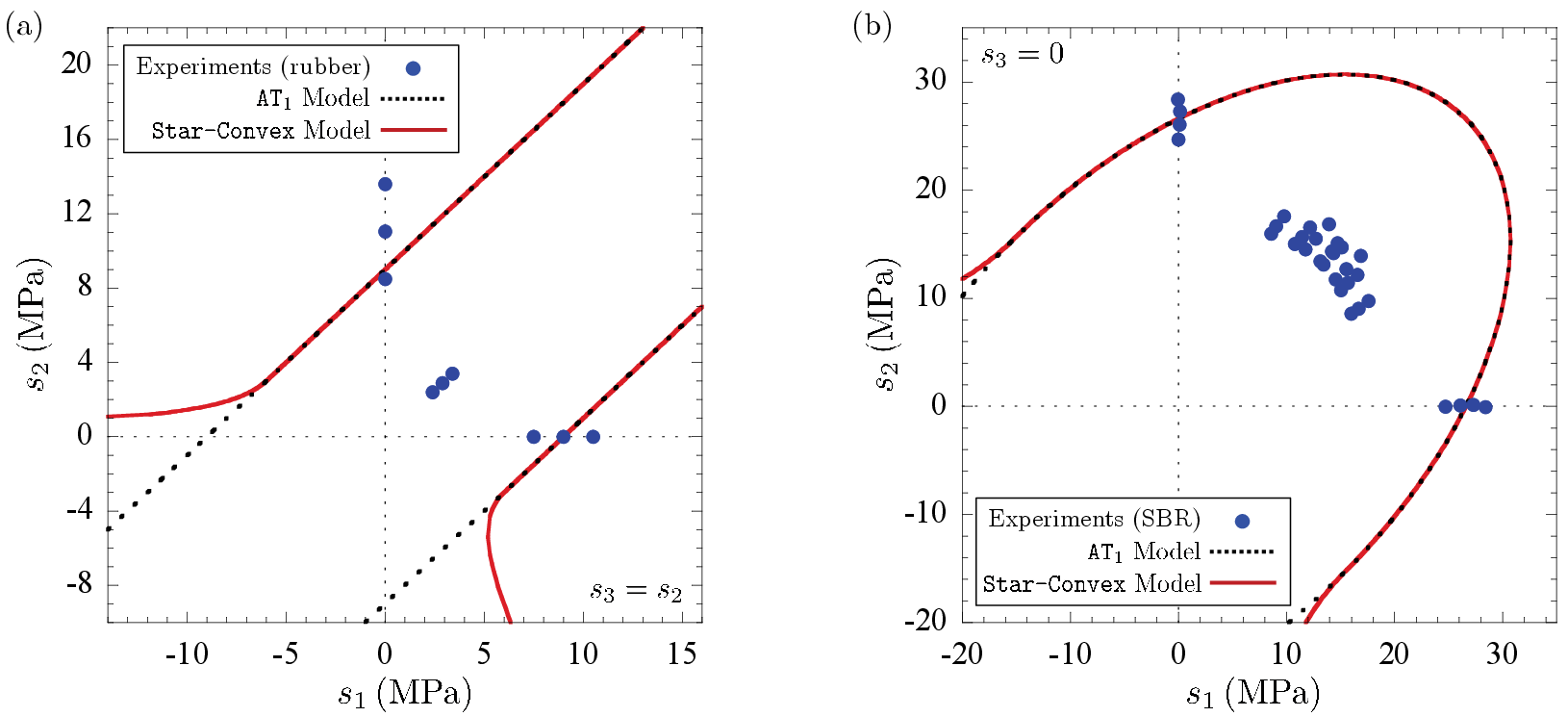}
   \caption{\small Comparisons of the strength surfaces (\ref{F-AT1}) and (\ref{F-Star}) generated by the \texttt{AT}$_1$ and \texttt{star-convex} variational phase-field models with experimental data for (a) natural rubber and (b) SBR. The data in part (a) is plotted in the space of principal stresses $(s_1, s_2)$ with $s_3 = s_2$, while the data  in part (b) is plotted in the space of principal stresses $(s_1, s_2)$ with $s_3 = 0$.}
   \label{Fig6}
\end{figure}

Next, Fig.~\ref{Fig6} confronts the strength surfaces (\ref{F-AT1}) and (\ref{F-Star}) generated by the \texttt{AT}$_1$ and \texttt{star-convex} variational phase-field models with experimental data for two elastomers of common use, natural rubber and a synthetic SBR rubber, studied respectively by \cite{GL59} and \cite{Hamdi06}. The data for natural rubber is plotted in the space of principal stresses $(s_1, s_2)$ with $s_3 = s_2$, while that for SBR is plotted in the space $(s_1, s_2)$ with $s_3 = 0$. The relevant material constants for these two elastomers are listed in Table \ref{Table1}, which also lists the values of the parameters $\varepsilon$ and $\gamma^\star$ in the strength surfaces (\ref{F-AT1}) and (\ref{F-Star}), as obtained from the prescriptions (\ref{eps-AT1}) and (\ref{gamma-scs}).  

Note that, save for the experimental data points around hydrostatic tension ($s_1=s_2=s_3$) in Fig.~\ref{Fig6}(a) for natural rubber, the remaining experimental data points in Fig.~\ref{Fig6} are associated with nominal stresses that are attained when the specimens are finitely deformed. This should call for the use of finite-elasticity versions of the \texttt{AT}$_1$ and \texttt{star-convex} variational phase-field models; see, e.g., \cite{KFLP18}. The use of finite elasticity would not significantly change the resulting strength surfaces in the space of principal nominal stresses $(s_1,s_2,s_3)$ shown in Fig.~\ref{Fig6}.

The main observations from the comparisons presented in Fig.~\ref{Fig6} are as follows. First, qualitatively, the shapes of the strength surfaces  (\ref{F-AT1}) and (\ref{F-Star}) are different from those of the experimental data, strikingly so for natural rubber. Second, quantitatively, both strength surfaces  (\ref{F-AT1}) and (\ref{F-Star}) do not provide a good fit with both sets of experimental results. This difference is clearly illustrated by the values obtained from the measures (\ref{R-measures}), which are presented in Table \ref{Table2}. These show that there is a direction in which the strength surfaces (\ref{F-AT1}) and (\ref{F-Star}) are 110 times larger than the corresponding experimental data for natural rubber, while for SBR there is a direction in which (\ref{F-AT1}) and (\ref{F-Star}) are 1.88 times larger than the corresponding experimental data. The extremely large difference for the case of natural rubber is due to the fact that the ratio of bulk-to-shear moduli of this material is large, $\kappa/\mu=3741$. As discussed above, because of their energetic nature, the strength surfaces (\ref{F-AT1}) and (\ref{F-Star}) describe a hydrostatic strength $\shs$ that is proportional to $\sqrt{\kappa}$. Clearly, this is inconsistent with the actual behavior of the material. 

\newpage

Summing up, the representative comparisons presented in Figs. \ref{Fig5} and \ref{Fig6} with experiments for four materials of common use, two hard and two soft, clearly illustrate that variational phase-field models, without and with energy splits, cannot possibly describe fracture nucleation in elastic brittle materials in general. The comparisons have also illustrated that such models may be calibrated to \emph{de facto} account for small portions of the actual strength surface of materials. While this may allow apparent agreement with particular experiments, it should be clear that the resulting models will not be descriptive nor predictive for problems beyond the very specific types of problems (those involving only the small portion of strength surface accounted for) that they have been calibrated to.

\section{Final comments}\label{Sec: Final Comments}

As reviewed above, the vast body of experimental evidence gathered over the years forces any potentially successful candidate for a macroscopic theory of fracture nucleation in nominally elastic brittle materials to account for the elasticity, the strength, and the toughness of the material of interest. Since variational phase-field models, without and with energy splits, cannot account for one of these three independent macroscopic material properties --- namely, the strength --- they are not viable candidates. 

By the same token, any formulation that cannot account for an arbitrary strength surface $\mathcal{F}(\bfS)=0$ is also non viable. In this regard, we note that existing cohesive models of fracture, be they sharp or regularized \citep{Dugdale60,Barenblatt62,Bourdin08,Iurlano16,Wu17}, while they do predict that nucleation does not occur unless a strength surface criterion is violated, cannot account for an arbitrary strength surface. 

As recalled in the Introduction, a class of phase-field models introduced by \cite*{KFLP18} provides a formulation that can account for the elasticity, the strength, and the toughness of elastic brittle materials, whatever these material properties may be. To that effect, the models introduce a driving force $c_\texttt{e}$ in the evolution equation for the phase field $v$. That force must be such that the phase field $v$ localizes properly and that the nucleation from large pre-existing cracks follows the Griffith energy competition. \cite{KLP20} and \cite{KBFLP20} offered a blueprint for such a driving force and used that blueprint to derive a particular one in the case of a Drucker-Prager type strength surface (\ref{DP})$_1$. In a recent contribution, \cite{KKLP24} have derived a new constitutive prescription for $c_\texttt{e}$ that has additional advantages over the original one, chief among them that it is fully explicit.

In the basic setting considered in Section \ref{Sec: The P-F model}, that of an isotropic linear elastic brittle material occupying an open bounded domain $\Omega_0\subset \mathbb{R}^3$, with boundary $\partial\Omega_0$ and outward unit normal $\bfN$, the phase-field model introduced by \cite*{KFLP18} with the driving force $c_\texttt{e}$ introduced by \cite{KKLP24} yields the following system of governing equations at any given discrete time $t_k\in\{0=t_0,t_1,...,t_m,t_{m+1},...,t_M=T\}$: 
\begin{equation}\vspace{-0.2cm}
\left\{\begin{array}{ll}
{\rm Div}\left[(v_{k}^2+\eta_{\varepsilon})\dfrac{\partial W}{\partial \bfE}(\bfE(\bfu_{k}))\right]=\textbf{0},& \,\bfX\in\Omega_0\vspace{0.2cm}\\
\bfu_k(\bfX)=\overline{\bfu}(\bfX,t_k), & \, \bfX\in\partial\Omega_0^{\mathcal{D}}\vspace{0.2cm}\\
\left[(v_{k}^2+\eta_{\varepsilon})\dfrac{\partial W}{\partial\bfE}(\bfE(\bfu_k))\right]\bfN=\overline{\textbf{s}}(\bfX,t_k),& \, \bfX\in\partial\Omega_0^{\mathcal{N}}
\end{array}\right. \label{BVP-u-theory}\vspace{-0.1cm}
\end{equation}
and
\begin{equation}\vspace{-0.1cm}
\left\{\begin{array}{lll}
\varepsilon\, \delta^\varepsilon G_c \Delta v_k=\dfrac{8}{3}v_{k} W(\bfE(\bfu_k))+\dfrac{4}{3}c_\texttt{e}(\bfX,t_{k})-\dfrac{\delta^\varepsilon G_c}{2\varepsilon},&\, \mbox{ if } v_{k}(\bfX)< v_{k-1}(\bfX),& \bfX\in \Omega_0\vspace{0.2cm}\\
\varepsilon\, \delta^\varepsilon G_c \Delta v_k\geq\dfrac{8}{3}v_{k} W(\bfE(\bfu_k))+\dfrac{4}{3}c_\texttt{e}(\bfX,t_{k})-\dfrac{\delta^\varepsilon G_c}{2\varepsilon},&\, \mbox{ if } v_{k}(\bfX)=1\; \mbox{ or }\; v_{k}(\bfX)= v_{k-1}(\bfX)>0, & \bfX\in \Omega_0\vspace{0.2cm}\\
v_k(\bfX)=0,&\, \mbox{ if } v_{k-1}(\bfX)=0, & \bfX\in \Omega_0\vspace{0.2cm}\\
\nabla v_k\cdot\bfN=0,&  &\, \bfX\in \partial\Omega_0
\end{array}\right.  \label{BVP-v-theory}
\end{equation}

\noindent with initial conditions $\bfu(\bfX,0)\equiv \textbf{0}$ and $v(\bfX,0)\equiv1$, where $\nabla v_k=\nabla v(\bfX,t_k)$, $\Delta v_k=\Delta v(\bfX,t_k)$,  $\eta_{\varepsilon}=o(\varepsilon)$ is a positive constant, and where
\begin{align}
c_{\texttt{e}}(\bfX,t)=\beta_2\sqrt{\mathcal{J}_2}+
\beta_1 \mathcal{I}_1-v\left(1-\dfrac{\sqrt{\mathcal{I}_1^2}}{\mathcal{I}_1}\right)W(\bfE(\bfu)).\label{cehat}
\end{align}
In (\ref{BVP-v-theory}) and (\ref{cehat}),
\begin{align*}
\left\{\hspace{-0.1cm}\begin{array}{l}\mathcal{I}_1={\rm tr}\,\bfS=3\kappa v^2 {\rm tr}\,\bfE\vspace{0.2cm}\\
\mathcal{J}_2=\dfrac{1}{2}\,{\rm tr}\,\bfS^2_D=2\mu^2 v^4 \left(\,{\rm tr}\, \bfE^2-\dfrac{1}{3}({\rm tr}\, \bfE)^2\right)\end{array}\right.\hspace{-0.1cm},\quad \left\{\hspace{-0.1cm}\begin{array}{l}\beta_1=\dfrac{1}{\shs}\delta^\varepsilon\dfrac{G_c}{8\varepsilon}-\dfrac{2\mathcal{W}_{\texttt{hs}}}{3\shs}\vspace{0.2cm}\\
\beta_2=\dfrac{\sqrt{3}(3\shs-\sts)}{\shs\sts}\delta^\varepsilon\dfrac{G_c}{8\varepsilon}+
\dfrac{2\mathcal{W}_{\texttt{hs}}}{\sqrt{3}\shs}-\dfrac{2\sqrt{3}\mathcal{W}_{\texttt{ts}}}{\sts}\end{array}\right.,
\end{align*}
and
\begin{align}
\delta^\varepsilon=\left(\dfrac{\sts+(1+2\sqrt{3})\,\shs}{(8+3\sqrt{3})\,\shs}\right)\dfrac{3 G_c}{16\mathcal{W}_{\texttt{ts}}\varepsilon}+\dfrac{2}{5},\qquad \mathcal{W}_{\texttt{ts}}=\dfrac{\sts^2}{2E}, \qquad \mathcal{W}_{\texttt{hs}}=\dfrac{\shs^2}{2\kappa}.\label{Wts}
\end{align}

The interested reader is referred to the above cited papers for a complete discussion of the theoretical and practical features of the phase-field model (\ref{BVP-u-theory})-(\ref{BVP-v-theory}) with (\ref{cehat}). Here, we wish to make three remarks of practical relevance.

First, on their own, the governing equations (\ref{BVP-u-theory}) and (\ref{BVP-v-theory}) are standard second-order PDEs for the displacement field $\bfu_k(\bfX)$ and the phase field $v_k(\bfX)$. Accordingly, their numerical solution is amenable to a FE staggered scheme in which (\ref{BVP-u-theory}) and (\ref{BVP-v-theory}) are discretized with finite elements and solved iteratively one after the other at every time step $t_k$ until convergence is reached. In a recent contribution, \cite{LDLP24} have shown that the solution pair $(\bfu^{\varepsilon}_k,v^{\varepsilon}_k)$ computed in such a staggered approach corresponds in fact to the fields that minimize separately two different functionals. This is exactly the same alternating minimization approach used to generate FE solutions for the classical variational phase-field models (\ref{BFM00}) and (\ref{W-Split}). The key difference lies in the accounting of the strength surface $\mathcal{F}(\bfS)=0$ of the material by (\ref{BVP-u-theory})-(\ref{BVP-v-theory}). 

Second, the phase-field model (\ref{BVP-u-theory})-(\ref{BVP-v-theory}) reduces identically to the \texttt{AT}$_1$ variational phase-field model (\ref{BFM00}) with (\ref{AT1AT2})$_1$ by setting $c_{\texttt{e}}(\bfX,t)=0$ and $\delta^\varepsilon=1$. This implies that any existing FE implementation of the \texttt{AT}$_1$ variational phase-field model can be readily upgraded to the phase-field model (\ref{BVP-u-theory})-(\ref{BVP-v-theory}) by simply multiplying $G_c$  with  (\ref{Wts})$_1$ and by adding the driving force (\ref{cehat}) to the right-hand side of the Euler-Largange equation associated with variations in the phase field $v$. FEniCSx and MOOSE implementations of the phase-field model (\ref{BVP-u-theory})-(\ref{BVP-v-theory}) are available on GitHub.\footnote{\url{https://github.com/farhadkama/FEniCSx_Kamarei_Kumar_Lopez-Pamies}.}$^,$\footnote{\url{https://github.com/hugary1995/raccoon}.}

Third, over the past six years, a series of simulations \citep{KFLP18,KRLP18,KBFLP20,KLP20,KLP21,KRLP22,KLDLP24,KKLP24} for numerous materials, hard and soft, specimen geometries, and loading conditions have shown that the predictions generated by the phase-field model (\ref{BVP-u-theory})-(\ref{BVP-v-theory}), as well as by its finite-elasticity version, are in qualitative and quantitative agreement with experimental observations on where and when fracture nucleates and propagates in nominally elastic brittle materials at large. 

It thus appears that a correct way to incorporate the strength into a phase-field model consists in adding a driving force to the $v$-equation, one tailored to the given strength surface $\mathcal{F}(\bfS)=0$. What the corresponding sharp model ($\varepsilon\searrow 0$) should look like for this class of phase-field models, if it even exists, is at present an open question.

\begin{figure}[t!]
   \centering \includegraphics[width=0.85\linewidth]{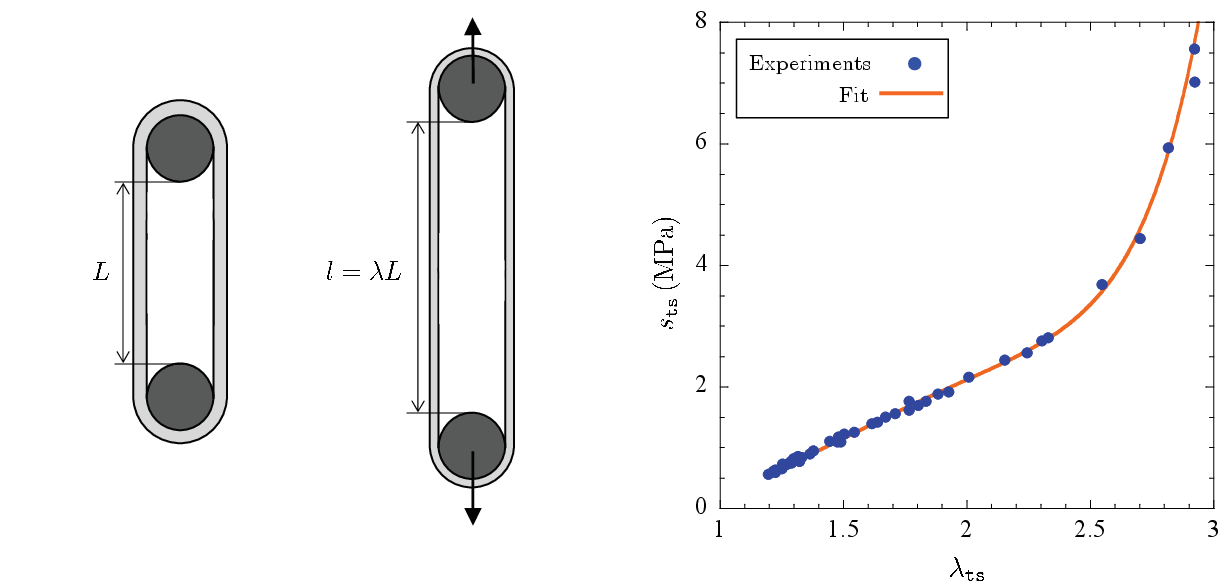}
   \caption{\small Classical experimental results due to \cite{Knauss67} for the uniaxial tensile strength $\sts$ of a polyurethane elastomer loaded at different constant quasi-static stretch rates. The data for $\sts$ is plotted as a function of the corresponding stretch $\lambda_{\texttt{ts}}$ at fracture. The schematic illustrates the type of ring specimens and loading setup used in the tests.}
   \label{Fig7}
\end{figure}

\medskip

What about the modeling of fracture nucleation beyond the basic setting of elastic brittle materials? 

\medskip

\noindent Actual materials are, of course, not purely elastic. They dissipate energy by deformation, typically via viscous and/or plastic processes, and not just by the creation of surface when fracturing. A review of experimental observations \citep{LP24}, akin to that presented in Section \ref{Sec: Experiments} above, reveals that once again fracture nucleation in dissipative materials critically depend on: (I) the property that describes the mechanics of deformation, (II) the property that describes the mechanics of strength, and (III) the toughness. This suggests that the same type of formulations may apply to materials at large and hence that such formulations may lead to a universal macroscopic theory of fracture. This is an exciting prospect.

In this regard, we note that the loading-history dependence of the mechanics of deformation of dissipative materials has long been a subject of intense activity in the mechanics community. The same is not true for the loading-history dependence of the mechanics of strength or of toughness. 

By way of an example, Fig.~\ref{Fig7} presents a set of experimental results due to \cite{Knauss67} of fracture nucleation in bands, also referred to as ring specimens, made of a polyurethane elastomer subjected to spatially uniform uniaxial tension applied at different constant quasi-static stretch rates. The main observation from these experiments is that larger stretch rates lead not only to larger values of the uniaxial tensile strength $\sts$ at which the specimens fracture, but also to larger values of the corresponding stretch $\lambda_{\texttt{ts}}$ at fracture.  This makes it plain that the strength of materials that dissipate energy by deformation --- in this case, by viscous deformation --- cannot be simply characterized by a fixed strength surface $\mathcal{F}(\bfS)=0$ in stress space. Instead, a more complex characterization, one that is loading-history dependent, is required. More studies in this direction would prove invaluable.

\section*{Acknowledgements}

O.L.P., J.E.D, and C.J.L. gratefully acknowledge support from the National Science Foundation through the Grants DMS--2308169,  CMMI--2132528, CMMI--2132551, and DMS--2206114.

\bibliographystyle{elsarticle-harv}
\bibliography{References}

\begin{thebibliography}{83}
\expandafter\ifx\csname natexlab\endcsname\relax\def\natexlab#1{#1}\fi
\providecommand{\url}[1]{\texttt{#1}}
\providecommand{\href}[2]{#2}
\providecommand{\path}[1]{#1}
\providecommand{\DOIprefix}{doi:}
\providecommand{\ArXivprefix}{arXiv:}
\providecommand{\URLprefix}{URL: }
\providecommand{\Pubmedprefix}{pmid:}
\providecommand{\doi}[1]{\href{http://dx.doi.org/#1}{\path{#1}}}
\providecommand{\Pubmed}[1]{\href{pmid:#1}{\path{#1}}}
\providecommand{\bibinfo}[2]{#2}
\ifx\xfnm\relax \def\xfnm[#1]{\unskip,\space#1}\fi
\bibitem[{Ambrosio and Tortorelli(1990)}]{AT90}
\bibinfo{author}{Ambrosio, L.}, \bibinfo{author}{Tortorelli, V.M.},
  \bibinfo{year}{1990}.
\newblock \bibinfo{title}{Approximation of functionals depending on jumps by
  elliptic functionals via {$\Gamma$}-convergence}.
\newblock \bibinfo{journal}{Communications on Pure and Applied Mathematics}
  \bibinfo{volume}{43}, \bibinfo{pages}{999--1036}.
\bibitem[{Ambrosio and Tortorelli(1992)}]{AT92}
\bibinfo{author}{Ambrosio, L.}, \bibinfo{author}{Tortorelli, V.M.},
  \bibinfo{year}{1992}.
\newblock \bibinfo{title}{On the approximation of free discontinuity problems}.
\newblock \bibinfo{journal}{Boll. Un. Mater. Ital. B} \bibinfo{volume}{6},
  \bibinfo{pages}{105--123}.
\bibitem[{Amor et~al.(2009)Amor, Marigo and Maurini}]{Amor09}
\bibinfo{author}{Amor, H.}, \bibinfo{author}{Marigo, J.J.},
  \bibinfo{author}{Maurini, C.}, \bibinfo{year}{2009}.
\newblock \bibinfo{title}{Regularized formulation of the variational brittle
  fracture with unilateral contact: {N}umerical experiments}.
\newblock \bibinfo{journal}{J. Mech. Phys. Solids} \bibinfo{volume}{57},
  \bibinfo{pages}{1209--1229}.
\bibitem[{Andrews(1963)}]{Andrews63}
\bibinfo{author}{Andrews, E.H.}, \bibinfo{year}{1963}.
\newblock \bibinfo{title}{Rupture propagation in hysteresial materials:
  {S}tress at a notch}.
\newblock \bibinfo{journal}{Journal of the Mechanics and Physics of Solids}
  \bibinfo{volume}{11}, \bibinfo{pages}{231--242}.
\bibitem[{Arunachala et~al.(2024)Arunachala, Vajari, Neuner, Sim, Zhao and
  Linder}]{Linder24}
\bibinfo{author}{Arunachala, P.K.}, \bibinfo{author}{Vajari, S.A.},
  \bibinfo{author}{Neuner, M.}, \bibinfo{author}{Sim, J.S.},
  \bibinfo{author}{Zhao, R.}, \bibinfo{author}{Linder, C.},
  \bibinfo{year}{2024}.
\newblock \bibinfo{title}{A multiscale anisotropic polymer network model
  coupled with phase field fracture}.
\newblock \bibinfo{journal}{Int. J. Numer. Methods Eng.} \bibinfo{volume}{125},
  \bibinfo{pages}{7488}.
\bibitem[{Awaji and Sato(1979)}]{Sato79}
\bibinfo{author}{Awaji, H.}, \bibinfo{author}{Sato, S.}, \bibinfo{year}{1979}.
\newblock \bibinfo{title}{Diametral compressive testing method}.
\newblock \bibinfo{journal}{J. Eng. Mater. Technol.} \bibinfo{volume}{101},
  \bibinfo{pages}{139--147}.
\bibitem[{Ball(1982)}]{Ball82}
\bibinfo{author}{Ball, J.M.}, \bibinfo{year}{1982}.
\newblock \bibinfo{title}{Discontinuous equilibrium solutions and cavitation in
  nonlinear elasticity}.
\newblock \bibinfo{journal}{Philos. Trans. R. Soc. A} \bibinfo{volume}{306},
  \bibinfo{pages}{557--611}.
\bibitem[{Barenblatt(1962)}]{Barenblatt62}
\bibinfo{author}{Barenblatt, G.I.}, \bibinfo{year}{1962}.
\newblock \bibinfo{title}{The mathematical theory of equilibrium cracks in
  brittle fracture}.
\newblock \bibinfo{journal}{Adv. Appl. Mech.} \bibinfo{volume}{7},
  \bibinfo{pages}{55--129}.
\bibitem[{Bharali et~al.(2022)Bharali, Goswami, Anitescu and
  Rabczuk}]{Rabczuk22}
\bibinfo{author}{Bharali, R.}, \bibinfo{author}{Goswami, S.},
  \bibinfo{author}{Anitescu, C.}, \bibinfo{author}{Rabczuk, T.},
  \bibinfo{year}{2022}.
\newblock \bibinfo{title}{A robust monolithic solver for phase-field fracture
  integrated with fracture energy based arc-length method and
  under-relaxation}.
\newblock \bibinfo{journal}{Computer Methods in Applied Mechanics and
  Engineering} \bibinfo{volume}{394}, \bibinfo{pages}{114927}.
\bibitem[{Bilgen and Weinberg(2019)}]{Weinberg19}
\bibinfo{author}{Bilgen, C.}, \bibinfo{author}{Weinberg, K.},
  \bibinfo{year}{2019}.
\newblock \bibinfo{title}{On the crack-driving force of phase-field models in
  linearized and finite elasticity}.
\newblock \bibinfo{journal}{Computer Methods in Applied Mechanics and
  Engineering} \bibinfo{volume}{253}, \bibinfo{pages}{348--372}.
\bibitem[{Bisai and Chakraborty(2019)}]{Bisai19}
\bibinfo{author}{Bisai, R.}, \bibinfo{author}{Chakraborty, S.},
  \bibinfo{year}{2019}.
\newblock \bibinfo{title}{Different failure modes of sandstone and shale under
  {B}razilian tensile tests}.
\newblock \bibinfo{journal}{Journal of Advances in Geotechnical Engineering}
  \bibinfo{volume}{2}, \bibinfo{pages}{1--8}.
\bibitem[{Borden et~al.(2012)Borden, Verhoosel, Scott, Hughes and
  Landis}]{Landis12}
\bibinfo{author}{Borden, M.J.}, \bibinfo{author}{Verhoosel, C.V.},
  \bibinfo{author}{Scott, M.A.}, \bibinfo{author}{Hughes, T.J.R.},
  \bibinfo{author}{Landis, C.M.}, \bibinfo{year}{2012}.
\newblock \bibinfo{title}{A phase-field description of dynamic brittle
  fracture}.
\newblock \bibinfo{journal}{Comput. Methods Appl. Mech. Engrg.}
  \bibinfo{volume}{217}, \bibinfo{pages}{77--95}.
\bibitem[{Bourdin et~al.(2000)Bourdin, Francfort and Marigo}]{Bourdin00}
\bibinfo{author}{Bourdin, B.}, \bibinfo{author}{Francfort, G.A.},
  \bibinfo{author}{Marigo, J.J.}, \bibinfo{year}{2000}.
\newblock \bibinfo{title}{Numerical experiments in revisited brittle fracture}.
\newblock \bibinfo{journal}{Journal of the Mechanics and Physics of Solids}
  \bibinfo{volume}{48}, \bibinfo{pages}{797--826}.
\bibitem[{Bourdin et~al.(2008)Bourdin, Francfort and Marigo}]{Bourdin08}
\bibinfo{author}{Bourdin, B.}, \bibinfo{author}{Francfort, G.A.},
  \bibinfo{author}{Marigo, J.J.}, \bibinfo{year}{2008}.
\newblock \bibinfo{title}{The variational approach to fracture}.
\newblock \bibinfo{journal}{Journal of Elasticity} \bibinfo{volume}{91},
  \bibinfo{pages}{5--148}.
\bibitem[{Braides(1998)}]{Braides98}
\bibinfo{author}{Braides, A.}, \bibinfo{year}{1998}.
\newblock \bibinfo{title}{Approximation of free-discontinuity problems}.
\newblock \bibinfo{publisher}{Springer}, \bibinfo{address}{Berlin}.
\bibitem[{Breedlove et~al.(2024)Breedlove, Chen, Lindeman and
  Lopez-Pamies}]{BCLLP24}
\bibinfo{author}{Breedlove, E.}, \bibinfo{author}{Chen, C.},
  \bibinfo{author}{Lindeman, D.}, \bibinfo{author}{Lopez-Pamies, O.},
  \bibinfo{year}{2024}.
\newblock \bibinfo{title}{Cavitation in elastomers: {A} review of the evidence
  against elasticity}.
\newblock \bibinfo{journal}{Journal of the Mechancis and Physics of Solids}
  \bibinfo{volume}{188}, \bibinfo{pages}{105678}.
\bibitem[{Busse(1934)}]{Busse34}
\bibinfo{author}{Busse, W.F.}, \bibinfo{year}{1934}.
\newblock \bibinfo{title}{Tear resistance and structure of rubber}.
\newblock \bibinfo{journal}{Ind. Eng. Chem.} \bibinfo{volume}{26},
  \bibinfo{pages}{1194--1199}.
\bibitem[{Chambolle et~al.(2018)Chambolle, Conti and Francfort}]{CCF18}
\bibinfo{author}{Chambolle, A.}, \bibinfo{author}{Conti, S.},
  \bibinfo{author}{Francfort, G.A.}, \bibinfo{year}{2018}.
\newblock \bibinfo{title}{Approximation of a brittle fracture energy with a
  constraint of non-interpenetration}.
\newblock \bibinfo{journal}{Arch. Rational Mech. Anal.} \bibinfo{volume}{228},
  \bibinfo{pages}{867--889}.
\bibitem[{Chen et~al.(2017)Chen, Wang and Suo}]{Chen17}
\bibinfo{author}{Chen, C.}, \bibinfo{author}{Wang, Z.}, \bibinfo{author}{Suo,
  Z.}, \bibinfo{year}{2017}.
\newblock \bibinfo{title}{Flaw sensitivity of highly stretchable materials}.
\newblock \bibinfo{journal}{Extreme Mechanics Letters} \bibinfo{volume}{10},
  \bibinfo{pages}{50--57}.
\bibitem[{Chen and Shen(2023)}]{Shen23}
\bibinfo{author}{Chen, Y.}, \bibinfo{author}{Shen, Y.}, \bibinfo{year}{2023}.
\newblock \bibinfo{title}{A ``parallel universe'' scheme for crack nucleation
  in the phase field approach to fracture}.
\newblock \bibinfo{journal}{Computer Methods in Applied Mechanics and
  Engineering} \bibinfo{volume}{403}, \bibinfo{pages}{115708}.
\bibitem[{Conti et~al.(2016)Conti, Focardi and Iurlano}]{Iurlano16}
\bibinfo{author}{Conti, S.}, \bibinfo{author}{Focardi, M.},
  \bibinfo{author}{Iurlano, F.}, \bibinfo{year}{2016}.
\newblock \bibinfo{title}{Phase field approximation of cohesive fracture
  models}.
\newblock \bibinfo{journal}{Ann. I. H. Poincar\'e -- AN} \bibinfo{volume}{33},
  \bibinfo{pages}{1033--1067}.
\bibitem[{{De Lorenzis} and Maurini(2022)}]{Maurini22}
\bibinfo{author}{{De Lorenzis}, L.}, \bibinfo{author}{Maurini, C.},
  \bibinfo{year}{2022}.
\newblock \bibinfo{title}{Nucleation under multi-axial loading in variational
  phase-field models of brittle fracture}.
\newblock \bibinfo{journal}{International Journal of Fracture}
  \bibinfo{volume}{237}, \bibinfo{pages}{61--81}.
\bibitem[{Dugdale(1960)}]{Dugdale60}
\bibinfo{author}{Dugdale, D.S.}, \bibinfo{year}{1960}.
\newblock \bibinfo{title}{Yielding of steel sheets containing slits}.
\newblock \bibinfo{journal}{Journal of the Mechanics and Physics of Solids}
  \bibinfo{volume}{8}, \bibinfo{pages}{100--104}.
\bibitem[{Dunn et~al.(1997)Dunn, Suwito and Cunningham}]{Dunn97}
\bibinfo{author}{Dunn, M.L.}, \bibinfo{author}{Suwito, W.},
  \bibinfo{author}{Cunningham, S.}, \bibinfo{year}{1997}.
\newblock \bibinfo{title}{Fracture initiation at sharp notches: {C}orrelation
  using critical stress intensities}.
\newblock \bibinfo{journal}{Int. J. Solids Struct.} \bibinfo{volume}{34},
  \bibinfo{pages}{3873--3883}.
\bibitem[{Ely(1972)}]{Ely72}
\bibinfo{author}{Ely, R.E.}, \bibinfo{year}{1972}.
\newblock \bibinfo{title}{Strength of titania and aluminum silicate under
  combined stresses}.
\newblock \bibinfo{journal}{J. Am. Ceram. Soc.} \bibinfo{volume}{55},
  \bibinfo{pages}{347--350}.
\bibitem[{Euchler et~al.(2020)Euchler, Bernhardt, Schneider, Heinrich,
  Wie{\ss}ner and Tada}]{Euchler20}
\bibinfo{author}{Euchler, E.}, \bibinfo{author}{Bernhardt, R.},
  \bibinfo{author}{Schneider, K.}, \bibinfo{author}{Heinrich, G.},
  \bibinfo{author}{Wie{\ss}ner, S.}, \bibinfo{author}{Tada, T.},
  \bibinfo{year}{2020}.
\newblock \bibinfo{title}{In situ dilatometry and {X}-ray microtomography study
  on the formation and growth of cavities in unfilled styrene-butadiene rubber
  vulcanizates subjected to constrained tensile deformation}.
\newblock \bibinfo{journal}{Polymer} \bibinfo{volume}{187},
  \bibinfo{pages}{122086}.
\bibitem[{Fan et~al.(2022)Fan, Jin and Wick}]{Wick22}
\bibinfo{author}{Fan, M.}, \bibinfo{author}{Jin, Y.}, \bibinfo{author}{Wick,
  T.}, \bibinfo{year}{2022}.
\newblock \bibinfo{title}{A quasi-monolithic phase-field description for
  mixed-mode fracture using predictor–corrector mesh adaptivity}.
\newblock \bibinfo{journal}{Engineering with Computers} \bibinfo{volume}{38},
  \bibinfo{pages}{S2879--S2903}.
\bibitem[{Ferreira et~al.(2024)Ferreira, Marengo and Perego}]{Perego24}
\bibinfo{author}{Ferreira, A.R.}, \bibinfo{author}{Marengo, A.},
  \bibinfo{author}{Perego, U.}, \bibinfo{year}{2024}.
\newblock \bibinfo{title}{A phase-field gradient-based energy split for the
  modeling of brittle fracture under load reversal}.
\newblock \bibinfo{journal}{Computer Methods in Applied Mechanics and
  Engineering} \bibinfo{volume}{431}, \bibinfo{pages}{117328}.
\bibitem[{Francfort and Marigo(1998)}]{Francfort98}
\bibinfo{author}{Francfort, G.A.}, \bibinfo{author}{Marigo, J.J.},
  \bibinfo{year}{1998}.
\newblock \bibinfo{title}{Revisiting brittle fracture as an energy minimization
  problem}.
\newblock \bibinfo{journal}{Journal of the Mechanics and Physics of Solids}
  \bibinfo{volume}{46}, \bibinfo{pages}{1319--1342}.
\bibitem[{Freddi and Carfagni(2010)}]{Freddi10}
\bibinfo{author}{Freddi, F.}, \bibinfo{author}{Carfagni, G.R.},
  \bibinfo{year}{2010}.
\newblock \bibinfo{title}{Regularized variational theories of fracture: {A}
  unified approach}.
\newblock \bibinfo{journal}{J. Mech. Phys. Solids} \bibinfo{volume}{58},
  \bibinfo{pages}{1154--1174}.
\bibitem[{Gent and Lindley(1959)}]{GL59}
\bibinfo{author}{Gent, A.N.}, \bibinfo{author}{Lindley, P.B.},
  \bibinfo{year}{1959}.
\newblock \bibinfo{title}{Internal rupture of bonded rubber cylinders in
  tension}.
\newblock \bibinfo{journal}{Proc. R. Soc. Lond. A} \bibinfo{volume}{249},
  \bibinfo{pages}{195--205}.
\bibitem[{Gent and Park(1984)}]{GentPark84}
\bibinfo{author}{Gent, A.N.}, \bibinfo{author}{Park, B.}, \bibinfo{year}{1984}.
\newblock \bibinfo{title}{Failure processes in elastomers at or near a rigid
  inclusion}.
\newblock \bibinfo{journal}{J. Mater. Sci.} \bibinfo{volume}{19},
  \bibinfo{pages}{1947--1956}.
\bibitem[{Giacomini and Ponsiglione(2008)}]{GP08}
\bibinfo{author}{Giacomini, A.}, \bibinfo{author}{Ponsiglione, M.},
  \bibinfo{year}{2008}.
\newblock \bibinfo{title}{Non interpenetration of matter for {SBV}-deformations
  of hyperelastic brittle materials}.
\newblock \bibinfo{journal}{Proc. R. Soc. Lond. A} \bibinfo{volume}{138A},
  \bibinfo{pages}{1019--1041}.
\bibitem[{Gomez et~al.(2005)Gomez, Elices and Planas}]{Gomez05}
\bibinfo{author}{Gomez, F.J.}, \bibinfo{author}{Elices, M.},
  \bibinfo{author}{Planas, J.}, \bibinfo{year}{2005}.
\newblock \bibinfo{title}{The cohesive crack concept: {A}pplications to {PMMA}
  at -$60$ $^\circ${C}}.
\newblock \bibinfo{journal}{Eng. Frac. Mech.} \bibinfo{volume}{72},
  \bibinfo{pages}{1268--1285}.
\bibitem[{Greensmith(1960)}]{Greensmith60}
\bibinfo{author}{Greensmith, H.W.}, \bibinfo{year}{1960}.
\newblock \bibinfo{title}{Rupture of rubber. {VIII}. {C}omparisons of tear and
  tensile rupture measurements}.
\newblock \bibinfo{journal}{J. Appl. Pol. Sci.} \bibinfo{volume}{3},
  \bibinfo{pages}{183--193}.
\bibitem[{Griffith(1921)}]{Griffith21}
\bibinfo{author}{Griffith, A.A.}, \bibinfo{year}{1921}.
\newblock \bibinfo{title}{The phenomena of rupture and flow in solids}.
\newblock \bibinfo{journal}{Philos. Trans. R. Soc. Lond. Ser. A}
  \bibinfo{volume}{221}, \bibinfo{pages}{163--198}.
\bibitem[{Guo and Ravi-Chandar(2023)}]{GuoRavi23}
\bibinfo{author}{Guo, J.}, \bibinfo{author}{Ravi-Chandar, K.},
  \bibinfo{year}{2023}.
\newblock \bibinfo{title}{On crack nucleation and propagation in elastomers:
  {I}. {I}n situ optical and {X}-ray experimental observations}.
\newblock \bibinfo{journal}{International Journal of Fracture}
  \bibinfo{volume}{243}, \bibinfo{pages}{1--29}.
\bibitem[{Hamdi et~al.(2006)Hamdi, {Nait Abdelaziz}, {Ait Hocine}, Heuillet and
  Benseddiq}]{Hamdi06}
\bibinfo{author}{Hamdi, A.}, \bibinfo{author}{{Nait Abdelaziz}, M.},
  \bibinfo{author}{{Ait Hocine}, N.}, \bibinfo{author}{Heuillet, P.},
  \bibinfo{author}{Benseddiq, N.}, \bibinfo{year}{2006}.
\newblock \bibinfo{title}{A fracture criterion of rubber-like materials under
  plane stress conditions}.
\newblock \bibinfo{journal}{Polym. Test.} \bibinfo{volume}{25},
  \bibinfo{pages}{994--1005}.
\bibitem[{Kamarei et~al.(2024)Kamarei, Kumar and Lopez-Pamies}]{KKLP24}
\bibinfo{author}{Kamarei, F.}, \bibinfo{author}{Kumar, A.},
  \bibinfo{author}{Lopez-Pamies, O.}, \bibinfo{year}{2024}.
\newblock \bibinfo{title}{The poker-chip experiments of synthetic elastomers
  explained}.
\newblock \bibinfo{journal}{Journal of the Mechanics and Physics of Solids}
  \bibinfo{volume}{188}, \bibinfo{pages}{105683}.
\bibitem[{Kasirajan et~al.(2020)Kasirajan, Bhattacharya, Rajagopal and
  Reddy}]{Reddy20}
\bibinfo{author}{Kasirajan, P.}, \bibinfo{author}{Bhattacharya, S.},
  \bibinfo{author}{Rajagopal, A.}, \bibinfo{author}{Reddy, J.N.},
  \bibinfo{year}{2020}.
\newblock \bibinfo{title}{Phase field modeling of fracture in quasi-brittle
  materials using natural neighbor {G}alerkin method}.
\newblock \bibinfo{journal}{Computer Methods in Applied Mechanics and
  Engineering} \bibinfo{volume}{366}, \bibinfo{pages}{113019}.
\bibitem[{Kawabata(1973)}]{Kawabata73}
\bibinfo{author}{Kawabata, S.}, \bibinfo{year}{1973}.
\newblock \bibinfo{title}{Fracture and mechanical behavior of rubber-like
  polymers under finite deformation in biaxial stress field}.
\newblock \bibinfo{journal}{J. Macromol. Sci. Part B} \bibinfo{volume}{8},
  \bibinfo{pages}{605--630}.
\bibitem[{Kimoto et~al.(1985)Kimoto, Usami and Miyata}]{Kimoto85}
\bibinfo{author}{Kimoto, H.}, \bibinfo{author}{Usami, S.},
  \bibinfo{author}{Miyata, H.}, \bibinfo{year}{1985}.
\newblock \bibinfo{title}{Flaw size dependence in fracture stress of glass and
  polycrystalline ceramics}.
\newblock \bibinfo{journal}{Transactions of the Japan Society of Mechanical
  Engineers Series A} \bibinfo{volume}{51}, \bibinfo{pages}{2482--2488}.
\bibitem[{Knauss(1967)}]{Knauss67}
\bibinfo{author}{Knauss, W.G.}, \bibinfo{year}{1967}.
\newblock \bibinfo{title}{An upper bound of failure in viscoelastic materials
  subjected to multiaxial stress states}.
\newblock \bibinfo{journal}{Int. J. Fract.} \bibinfo{volume}{3},
  \bibinfo{pages}{267--277}.
\bibitem[{Kovar et~al.(2011)Kovar, Bennison and Readey}]{Kovar00}
\bibinfo{author}{Kovar, D.}, \bibinfo{author}{Bennison, S.J.},
  \bibinfo{author}{Readey, M.J.}, \bibinfo{year}{2011}.
\newblock \bibinfo{title}{Crack stability and strength variability in alumina
  ceramics with rising toughness-curve behavior}.
\newblock \bibinfo{journal}{Acta Materialia} \bibinfo{volume}{48},
  \bibinfo{pages}{565--578}.
\bibitem[{Kruzic et~al.(2004)Kruzic, Cannon and Ritchie}]{Ritchie04}
\bibinfo{author}{Kruzic, J.J.}, \bibinfo{author}{Cannon, R.M.},
  \bibinfo{author}{Ritchie, R.O.}, \bibinfo{year}{2004}.
\newblock \bibinfo{title}{Crack-size effects on cyclic and monotonic crack
  growth in polycrystalline alumina: {Q}uantification of the role of grain
  bridging}.
\newblock \bibinfo{journal}{J. Am. Ceram. Sor.} \bibinfo{volume}{87},
  \bibinfo{pages}{93--103}.
\bibitem[{Kuhn and M\"uller(2010)}]{Muller10}
\bibinfo{author}{Kuhn, C.}, \bibinfo{author}{M\"uller, R.},
  \bibinfo{year}{2010}.
\newblock \bibinfo{title}{A continuum phase field model for fracture}.
\newblock \bibinfo{journal}{Eng. Fract. Mech.} \bibinfo{volume}{77},
  \bibinfo{pages}{3625--3634}.
\bibitem[{Kumar et~al.(2020)Kumar, Bourdin, Francfort and
  Lopez-Pamies}]{KBFLP20}
\bibinfo{author}{Kumar, A.}, \bibinfo{author}{Bourdin, B.},
  \bibinfo{author}{Francfort, G.A.}, \bibinfo{author}{Lopez-Pamies, O.},
  \bibinfo{year}{2020}.
\newblock \bibinfo{title}{Revisiting nucleation in the phase-field approach to
  brittle fracture}.
\newblock \bibinfo{journal}{Journal of the Mechanics and Physics of Solids}
  \bibinfo{volume}{142}, \bibinfo{pages}{104027}.
\bibitem[{Kumar et~al.(2018a)Kumar, Francfort and Lopez-Pamies}]{KFLP18}
\bibinfo{author}{Kumar, A.}, \bibinfo{author}{Francfort, G.A.},
  \bibinfo{author}{Lopez-Pamies, O.}, \bibinfo{year}{2018}a.
\newblock \bibinfo{title}{Fracture and healing of elastomers: {A}
  phase-transition theory and numerical implementation}.
\newblock \bibinfo{journal}{Journal of the Mechanics and Physics of Solids}
  \bibinfo{volume}{112}, \bibinfo{pages}{523--551}.
\bibitem[{Kumar et~al.(2024)Kumar, Liu, Dolbow and Lopez-Pamies}]{KLDLP24}
\bibinfo{author}{Kumar, A.}, \bibinfo{author}{Liu, Y.},
  \bibinfo{author}{Dolbow, J.E.}, \bibinfo{author}{Lopez-Pamies, O.},
  \bibinfo{year}{2024}.
\newblock \bibinfo{title}{The strength of the {B}razilian fracture test}.
\newblock \bibinfo{journal}{Journal of the Mechanics and Physics of Solids}
  \bibinfo{volume}{182}, \bibinfo{pages}{105473}.
\bibitem[{Kumar and Lopez-Pamies(2020)}]{KLP20}
\bibinfo{author}{Kumar, A.}, \bibinfo{author}{Lopez-Pamies, O.},
  \bibinfo{year}{2020}.
\newblock \bibinfo{title}{The phase-field approach to self-healable fracture of
  elastomers: {A} model accounting for fracture nucleation at large, with
  application to a class of conspicuous experiments}.
\newblock \bibinfo{journal}{Theoretical and Applied Fracture Mechanics}
  \bibinfo{volume}{107}, \bibinfo{pages}{102550}.
\bibitem[{Kumar and Lopez-Pamies(2021)}]{KLP21}
\bibinfo{author}{Kumar, A.}, \bibinfo{author}{Lopez-Pamies, O.},
  \bibinfo{year}{2021}.
\newblock \bibinfo{title}{The poker-chip experiments of {G}ent and {L}indley
  (1959) explained}.
\newblock \bibinfo{journal}{Journal of the Mechanics and Physics of Solids}
  \bibinfo{volume}{150}, \bibinfo{pages}{104359}.
\bibitem[{Kumar et~al.(2018b)Kumar, Ravi-Chandar and Lopez-Pamies}]{KRLP18}
\bibinfo{author}{Kumar, A.}, \bibinfo{author}{Ravi-Chandar, K.},
  \bibinfo{author}{Lopez-Pamies, O.}, \bibinfo{year}{2018}b.
\newblock \bibinfo{title}{The configurational-forces view of fracture and
  healing in elastomers as a phase transition}.
\newblock \bibinfo{journal}{International Journal of Fracture}
  \bibinfo{volume}{213}, \bibinfo{pages}{1--16}.
\bibitem[{Kumar et~al.(2022)Kumar, Ravi-Chandar and Lopez-Pamies}]{KRLP22}
\bibinfo{author}{Kumar, A.}, \bibinfo{author}{Ravi-Chandar, K.},
  \bibinfo{author}{Lopez-Pamies, O.}, \bibinfo{year}{2022}.
\newblock \bibinfo{title}{The revisited phase-field approach to brittle
  fracture: {A}pplication to indentation and notch problems}.
\newblock \bibinfo{journal}{International Journal of Fracture}
  \bibinfo{volume}{237}, \bibinfo{pages}{83--100}.
\bibitem[{Lam\'e and Clapeyron(1833)}]{Lame1833}
\bibinfo{author}{Lam\'e, G.}, \bibinfo{author}{Clapeyron, B.P.E.},
  \bibinfo{year}{1833}.
\newblock \bibinfo{title}{Memoire sur l'equilibre interieur des corps solides
  homogenes. [{M}emoir on the internal equilibrium of homogeneous solid
  bodies.]}.
\newblock \bibinfo{journal}{Paris, Mem. Par Divers Savants} ,
  \bibinfo{pages}{145--169}.
\bibitem[{Larsen(2021)}]{Larsen21}
\bibinfo{author}{Larsen, C.J.}, \bibinfo{year}{2021}.
\newblock \bibinfo{title}{Variational fracture with boundary loads}.
\newblock \bibinfo{journal}{Applied Mathematics Letters} \bibinfo{volume}{121},
  \bibinfo{pages}{107437}.
\bibitem[{Larsen et~al.(2024)Larsen, Dolbow and Lopez-Pamies}]{LDLP24}
\bibinfo{author}{Larsen, C.J.}, \bibinfo{author}{Dolbow, J.E.},
  \bibinfo{author}{Lopez-Pamies, O.}, \bibinfo{year}{2024}.
\newblock \bibinfo{title}{A variational formulation of {G}riffith phase-field
  fracture with material strength}.
\newblock \bibinfo{journal}{International Journal of Fracture}
  \bibinfo{volume}{247}, \bibinfo{pages}{319--327}.
\bibitem[{Lawn(1998)}]{Lawn98}
\bibinfo{author}{Lawn, B.R.}, \bibinfo{year}{1998}.
\newblock \bibinfo{title}{Indentation of ceramics with spheres: {A} century
  after {H}ertz}.
\newblock \bibinfo{journal}{J. Am. Ceram. Soc.} \bibinfo{volume}{81},
  \bibinfo{pages}{1977--1994}.
\bibitem[{Lef\`evre et~al.(2015)Lef\`evre, Ravi-Chandar and
  Lopez-Pamies}]{LRLP15}
\bibinfo{author}{Lef\`evre, V.}, \bibinfo{author}{Ravi-Chandar, K.},
  \bibinfo{author}{Lopez-Pamies, O.}, \bibinfo{year}{2015}.
\newblock \bibinfo{title}{Cavitation in rubber: {A}n elastic instability or a
  fracture phenomenon?}
\newblock \bibinfo{journal}{International Journal of Fracture}
  \bibinfo{volume}{192}, \bibinfo{pages}{1--23}.
\bibitem[{Li and Bouklas(2020)}]{Bouklas20}
\bibinfo{author}{Li, B.}, \bibinfo{author}{Bouklas, N.}, \bibinfo{year}{2020}.
\newblock \bibinfo{title}{A variational phase-field model for brittle fracture
  in polydisperse elastomer networks}.
\newblock \bibinfo{journal}{International Journal of Solids and Structures}
  \bibinfo{volume}{182}, \bibinfo{pages}{193--204}.
\bibitem[{Lopez-Pamies(2023)}]{LP23}
\bibinfo{author}{Lopez-Pamies, O.}, \bibinfo{year}{2023}.
\newblock \bibinfo{title}{Journal {C}lub: {S}trength revisited: {O}ne of three
  basic ingredients needed for a complete macroscopic theory of fracture}.
\newblock \URLprefix \url{https://imechanica.org/node/26641}.
\bibitem[{Lopez-Pamies(2024)}]{LP24}
\bibinfo{author}{Lopez-Pamies, O.}, \bibinfo{year}{2024}.
\newblock \bibinfo{title}{Nucleation and propagation of fracture in
  viscoelastic elastomers: {A} complete phase-field theory}.
\newblock \bibinfo{journal}{In preparation} .
\bibitem[{Lorentz et~al.(2011)Lorentz, Cuvilliez and Kazymyrenko}]{Lorentz11}
\bibinfo{author}{Lorentz, E.}, \bibinfo{author}{Cuvilliez, S.},
  \bibinfo{author}{Kazymyrenko, K.}, \bibinfo{year}{2011}.
\newblock \bibinfo{title}{Convergence of a gradient damage model toward a
  cohesive zone model}.
\newblock \bibinfo{journal}{Comptes Rendus Mecanique} \bibinfo{volume}{339},
  \bibinfo{pages}{20--26}.
\bibitem[{{Martinez-Pa\~neda} et~al.(2018){Martinez-Pa\~neda}, Golahmarb and
  Niordson}]{Paneda18}
\bibinfo{author}{{Martinez-Pa\~neda}, E.}, \bibinfo{author}{Golahmarb, A.},
  \bibinfo{author}{Niordson, C.F.}, \bibinfo{year}{2018}.
\newblock \bibinfo{title}{A phase field formulation for hydrogen assisted
  cracking}.
\newblock \bibinfo{journal}{Comput. Methods Appl. Mech. Engrg.}
  \bibinfo{volume}{342}, \bibinfo{pages}{742--761}.
\bibitem[{Miehe et~al.(2015)Miehe, Sch\"anzel and Ulmer}]{Miehe15}
\bibinfo{author}{Miehe, C.}, \bibinfo{author}{Sch\"anzel, L.M.},
  \bibinfo{author}{Ulmer, H.}, \bibinfo{year}{2015}.
\newblock \bibinfo{title}{Phase field modeling of fracture in multi-physics
  problems. {P}art {I}. {B}alance of crack surface and failure criteria for
  brittle crack propagation in thermo-elastic solids}.
\newblock \bibinfo{journal}{Computer Methods in Applied Mechanics and
  Engineering} \bibinfo{volume}{294}, \bibinfo{pages}{449--485}.
\bibitem[{Miehe et~al.(2010)Miehe, Welschinger and Hofacker}]{Miehe10}
\bibinfo{author}{Miehe, C.}, \bibinfo{author}{Welschinger, F.},
  \bibinfo{author}{Hofacker, M.}, \bibinfo{year}{2010}.
\newblock \bibinfo{title}{Thermodynamically consistent phase-field models of
  fracture: variational principles and multi-field {FE} implementations}.
\newblock \bibinfo{journal}{Int. J. Numer. Methods Eng.} \bibinfo{volume}{83},
  \bibinfo{pages}{1273--1311}.
\bibitem[{Mouginot and Maugis(1985)}]{Mouginot85}
\bibinfo{author}{Mouginot, R.}, \bibinfo{author}{Maugis, D.},
  \bibinfo{year}{1985}.
\newblock \bibinfo{title}{Fracture indentation beneath flat and spherical
  punches}.
\newblock \bibinfo{journal}{J. Mater. Sci.} \bibinfo{volume}{20},
  \bibinfo{pages}{4354--4376}.
\bibitem[{Noii et~al.(2020)Noii, Aldakheel, Wick and Wriggers}]{Wriggers20}
\bibinfo{author}{Noii, N.}, \bibinfo{author}{Aldakheel, F.},
  \bibinfo{author}{Wick, T.}, \bibinfo{author}{Wriggers, P.},
  \bibinfo{year}{2020}.
\newblock \bibinfo{title}{An adaptive global–local approach for phase-field
  modeling of anisotropic brittle fracture}.
\newblock \bibinfo{journal}{Computer Methods in Applied Mechanics and
  Engineering} \bibinfo{volume}{361}, \bibinfo{pages}{112744}.
\bibitem[{Pham et~al.(2011)Pham, Amor, Marigo and Maurini}]{Marigo11}
\bibinfo{author}{Pham, K.}, \bibinfo{author}{Amor, H.},
  \bibinfo{author}{Marigo, J.J.}, \bibinfo{author}{Maurini, C.},
  \bibinfo{year}{2011}.
\newblock \bibinfo{title}{Gradient damage models and their use to approximate
  brittle fracture}.
\newblock \bibinfo{journal}{Int. J. Damage Mech.} \bibinfo{volume}{20},
  \bibinfo{pages}{618--652}.
\bibitem[{Poulain et~al.(2017)Poulain, Lef\`evre, Lopez-Pamies and
  Ravi-Chandar}]{Poulain17}
\bibinfo{author}{Poulain, X.}, \bibinfo{author}{Lef\`evre, V.},
  \bibinfo{author}{Lopez-Pamies, O.}, \bibinfo{author}{Ravi-Chandar, K.},
  \bibinfo{year}{2017}.
\newblock \bibinfo{title}{Damage in elastomers: {N}ucleation and growth of
  cavities, micro-cracks, and macro-cracks}.
\newblock \bibinfo{journal}{International Journal of Fracture}
  \bibinfo{volume}{205}, \bibinfo{pages}{1--21}.
\bibitem[{Poulain et~al.(2018)Poulain, Lopez-Pamies and
  Ravi-Chandar}]{Poulain18}
\bibinfo{author}{Poulain, X.}, \bibinfo{author}{Lopez-Pamies, O.},
  \bibinfo{author}{Ravi-Chandar, K.}, \bibinfo{year}{2018}.
\newblock \bibinfo{title}{Damage in elastomers: {H}ealing of internally
  nucleated cavities and micro-cracks}.
\newblock \bibinfo{journal}{Soft Matter} \bibinfo{volume}{14},
  \bibinfo{pages}{4633--4640}.
\bibitem[{Rivlin and Thomas(1953)}]{RT53}
\bibinfo{author}{Rivlin, R.S.}, \bibinfo{author}{Thomas, A.G.},
  \bibinfo{year}{1953}.
\newblock \bibinfo{title}{Rupture of rubber. {I}. {C}haracteristic energy for
  tearing}.
\newblock \bibinfo{journal}{Journal of Polymer Science} \bibinfo{volume}{10},
  \bibinfo{pages}{291--318}.
\bibitem[{Roesler(1956)}]{Roesler56}
\bibinfo{author}{Roesler, F.C.}, \bibinfo{year}{1956}.
\newblock \bibinfo{title}{Brittle fractures near equilibrium}.
\newblock \bibinfo{journal}{Proceedings of the Physical Society. Section B}
  \bibinfo{volume}{69}, \bibinfo{pages}{981--992}.
\bibitem[{Sato et~al.(1987)Sato, Awaji, Kawamata, Kurumada and Oku}]{Sato87}
\bibinfo{author}{Sato, S.}, \bibinfo{author}{Awaji, H.},
  \bibinfo{author}{Kawamata, K.}, \bibinfo{author}{Kurumada, A.},
  \bibinfo{author}{Oku, T.}, \bibinfo{year}{1987}.
\newblock \bibinfo{title}{Fracture criteria of reactor graphite under
  multiaxial stresses}.
\newblock \bibinfo{journal}{Nuclear Engng. Design} \bibinfo{volume}{103},
  \bibinfo{pages}{291--300}.
\bibitem[{Sheikh et~al.(2019)Sheikh, Wang, Du, Suo, Li, Zhou, Wang, Dar, Gao
  and Wang}]{Sheikh19}
\bibinfo{author}{Sheikh, M.Z.}, \bibinfo{author}{Wang, Z.},
  \bibinfo{author}{Du, B.}, \bibinfo{author}{Suo, T.}, \bibinfo{author}{Li,
  Y.}, \bibinfo{author}{Zhou, F.}, \bibinfo{author}{Wang, Y.},
  \bibinfo{author}{Dar, U.A.}, \bibinfo{author}{Gao, G.},
  \bibinfo{author}{Wang, Y.}, \bibinfo{year}{2019}.
\newblock \bibinfo{title}{Static and dynamic {B}razilian disk tests for
  mechanical characterization of annealed and chemically strengthened glass}.
\newblock \bibinfo{journal}{Ceramics International} \bibinfo{volume}{45},
  \bibinfo{pages}{7931--7944}.
\bibitem[{Steinke and Kaliske(2019)}]{Kaliske19}
\bibinfo{author}{Steinke, C.}, \bibinfo{author}{Kaliske, M.},
  \bibinfo{year}{2019}.
\newblock \bibinfo{title}{A phase-field crack model based on directional stress
  decomposition}.
\newblock \bibinfo{journal}{Computational Mechanics} \bibinfo{volume}{63},
  \bibinfo{pages}{1019--1046}.
\bibitem[{Swamynathan et~al.(2021)Swamynathan, Jobst and Keip}]{Keip21}
\bibinfo{author}{Swamynathan, S.}, \bibinfo{author}{Jobst, S.},
  \bibinfo{author}{Keip, M.A.}, \bibinfo{year}{2021}.
\newblock \bibinfo{title}{An energetically consistent tension–compression
  split for phase-field models of fracture at large deformations}.
\newblock \bibinfo{journal}{Mechanics of Materials} \bibinfo{volume}{157},
  \bibinfo{pages}{103802}.
\bibitem[{Tada et~al.(1973)Tada, Paris and Irwin}]{Tada73}
\bibinfo{author}{Tada, H.}, \bibinfo{author}{Paris, P.C.},
  \bibinfo{author}{Irwin, G.R.}, \bibinfo{year}{1973}.
\newblock \bibinfo{title}{The stress analysis of cracks handbook 3rd edition}.
\newblock \bibinfo{publisher}{The American Society of Mechanical Engineers},
  \bibinfo{address}{New York}.
\bibitem[{Tang et~al.(2019)Tang, Zhang, Guo, Guo and Liu}]{WKLiu19}
\bibinfo{author}{Tang, S.}, \bibinfo{author}{Zhang, G.}, \bibinfo{author}{Guo,
  T.F.}, \bibinfo{author}{Guo, X.}, \bibinfo{author}{Liu, W.K.},
  \bibinfo{year}{2019}.
\newblock \bibinfo{title}{Phase field modeling of fracture in nonlinearly
  elastic solids via energy decomposition}.
\newblock \bibinfo{journal}{Computer Methods in Applied Mechanics and
  Engineering} \bibinfo{volume}{347}, \bibinfo{pages}{477--494}.
\bibitem[{Tann\'e et~al.(2018)Tann\'e, Li, Bourdin, Marigo and
  Maurini}]{Tanne18}
\bibinfo{author}{Tann\'e, E.}, \bibinfo{author}{Li, T.},
  \bibinfo{author}{Bourdin, B.}, \bibinfo{author}{Marigo, J.J.},
  \bibinfo{author}{Maurini, C.}, \bibinfo{year}{2018}.
\newblock \bibinfo{title}{Crack nucleation in variational phase-field models of
  brittle fracture}.
\newblock \bibinfo{journal}{J. Mech. Phys. Solids} \bibinfo{volume}{110},
  \bibinfo{pages}{80--99}.
\bibitem[{Thomas and Whittle(1970)}]{Thomas1970}
\bibinfo{author}{Thomas, A.G.}, \bibinfo{author}{Whittle, J.M.},
  \bibinfo{year}{1970}.
\newblock \bibinfo{title}{Tensile rupture of rubber}.
\newblock \bibinfo{journal}{Rubber Chemistry and Technology}
  \bibinfo{volume}{43}, \bibinfo{pages}{222--228}.
\bibitem[{Valentin et~al.(2010)Valentin, Posadas, Fernández-Torres, Malmierca,
  Gonzalez, Chasse and Saalw\"{a}chter}]{Valentinetal2010}
\bibinfo{author}{Valentin, J.L.}, \bibinfo{author}{Posadas, P.},
  \bibinfo{author}{Fernández-Torres, A.}, \bibinfo{author}{Malmierca, M.A.},
  \bibinfo{author}{Gonzalez, L.}, \bibinfo{author}{Chasse, W.},
  \bibinfo{author}{Saalw\"{a}chter, K.}, \bibinfo{year}{2010}.
\newblock \bibinfo{title}{Inhomogeneities and chain dynamics in diene rubbers
  vulcanized with different cure systems}.
\newblock \bibinfo{journal}{Macromolecules} \bibinfo{volume}{43},
  \bibinfo{pages}{4210--4222}.
\bibitem[{Vicentini et~al.(2024)Vicentini, Zolesi, Carrara, Maurini and {De
  Lorenzis}}]{Maurini24}
\bibinfo{author}{Vicentini, F.}, \bibinfo{author}{Zolesi, C.},
  \bibinfo{author}{Carrara, P.}, \bibinfo{author}{Maurini, C.},
  \bibinfo{author}{{De Lorenzis}, L.}, \bibinfo{year}{2024}.
\newblock \bibinfo{title}{On the energy decomposition in variational
  phase-field models for brittle fracture under multi-axial stress states}.
\newblock \bibinfo{journal}{International Journal of Fracture.}
  \bibinfo{volume}{In press}.
\bibitem[{Wu(2017)}]{Wu17}
\bibinfo{author}{Wu, J.Y.}, \bibinfo{year}{2017}.
\newblock \bibinfo{title}{A unified phase-field theory for the mechanics of
  damage and quasi-brittle failure}.
\newblock \bibinfo{journal}{Journal of the Mechanics and Physics of Solids}
  \bibinfo{volume}{103}, \bibinfo{pages}{72--99}.

\end{thebibliography}

\end{document}